\documentclass[prc,twocolumn]{revtex4-2}
\usepackage{color}

\usepackage{amsmath}
\usepackage{amssymb}
\usepackage{graphicx}
\usepackage{bm}

\renewcommand{\vec}[1]{\mbox{\boldmath $#1$}}

\begin{document}
\title{Generator coordinate method with a conjugate momentum: application to the
particle number projection}

\author{N. Hizawa}
\author{K. Hagino}
\author{K. Yoshida}
\affiliation{
Department of Physics, Kyoto University, Kyoto 606-8502,  Japan}

\begin{abstract}
We discuss an extension of the generator coordinate method (GCM)
by taking simultaneously a collective coordinate and
its conjugate momentum as generator coordinates.
To this end, we follow the idea of the dynamical GCM (DGCM) proposed
by Goeke and Reinhard.
We first show that the DGCM method can be regarded as an extension of
the double projection
method for the center of mass motion.
As an application of DGCM, we then investigate the particle number projection,
for which
we not only carry out an integral over the gauge angle as in the usual
particle number projection but
also take a linear superposition of BCS states which have different
mean particle numbers.
We show that the ground state energy is significantly lowered by such
effect, especially for magic nuclei for which the pairing gap is zero in the BCS
approximation.
This suggests that the present method makes a good alternative to the variation
after projection (VAP) method, as the method is much simpler than the VAP.
\end{abstract}

\maketitle

\section{Introduction}

Beyond mean-field calculations based on the generator coordinate method (GCM)
have been rapidly developing in recent years
\cite{Bender2003,NV11,E16,RR18,BH03,DB03,BF03,RE04,BH04,SO06,RE07,TR08,BH08,YM10,RE10,RT11,TR11,RE11,YB13,FS13,BA14,Yao2014,YZ15,EJ20}.
In this method, the wave function of many-body states is described as a linear
superposition of many Slater determinants \cite{Ring_Schuck}.
In this way,
quantum correlations beyond the mean
field approximation
are incorporated in the ground-state wave function.
Furthermore, the GCM provides not only the ground state but also excited states
described by the chosen generator coordinates.
For these reasons, the GCM has often been employed
for a microscopic description of nuclear collective motions.

While the idea of GCM is conceptually simple,
it has also long been known that a naive GCM cannot describe properly
the center of motion of a nucleus. That is, the appropriate moment of inertia
associated with the translational motion, i.e., the total mass of a system,
cannot be obtained
by simply superposing the wave
functions located at different center of mass positions \cite{Ring_Schuck}.
Peierls and Thouless resolved this problem by
projecting the GCM state on a state with a definite linear momentum \cite{PT62}.
This method has been referred to as the double projection method, which has also
been formulated for a rotational motion \cite{PT62}.

Recently, the idea based on the double projection method has been put forward
by
Borrajo {\it et al.} \cite{BR15,EB16} as well as
by Shimada {\it et al.} \cite{ST15,Shimada2016,Ushitani2019} for a calculation of
rotational bands in deformed nuclei.
These authors employed cranked-Hartree--Fock--Bogoliubov states with
several deformations and rotational frequencies as basis states for the GCM
calculations after
performing the angular momentum projection.
By including
the cranked states with a broken time-reversal symmetry, it was shown that
the excitation energies in the rotational bands are significantly lowered.

One can view these results form a different point of view.
That is,
the angular momentum projection is carried out
by superposing many-body states with different orientation angles of the principle
axes. To this end, the angular momentum projection is usually
applied to the states with a
single rotational frequency only.
Superposing cranked states with different rotational frequencies implies that
the quantity conjugate to the angle, that is, the angular momentum,
is incorporated in GCM states.
This suggests that
one can achieve a better description of collective states
by simultaneously treating a collective coordinate and its conjugate momentum
in the GCM method.

In addition to the double projection method,
there are several ways to extend the GCM along this line, such as
the complex GCM, in which collective coordinates
are regarded as complex numbers \cite{Ring_Schuck,jan64,BW68}.
In this connection,
we mention that Goeke and Reinhard have formulated the GCM by
introducing
the conjugate momentum apart from a collective variable
and called it the dynamical GCM (DGCM) \cite{GR78,RG78a,RG78b,RG78b,RG78c,GR80}.
It has been argued that the DGCM includes the complex GCM as a special case
and that the DGCM is an extension of the double projection method.
However, due to its complexity, no concrete numerical calculations have been
carried out with DGCM as far as the authors know.

The aim of this paper is to apply the DGCM
to the particle number fluctuation in a BCS wave function, for which
a pair of the canonical variables is known a priori, that is, the gauge angle and the
particle number.
Treating these variables as generator coordinates is nothing but
an application of the DGCM.
This amounts to superposing many BCS states with different
particle numbers after performing the particle number projection,
as in Refs. \cite{BR15,EB16,ST15,Shimada2016,Ushitani2019}
for rotational motions.
This work can in fact be regarded as the first step in a long-range project
of applications
of the DGCM to nuclear collective motions.
The fact that there is only a single variable (and its conjugate)
for the collective coordinate is another numerical advantage to investigating
the particle number fluctuation.

The paper is organized as follows.
In Sec. II, we give a brief review of the DGCM.
We then show that the DGCM is equivalent to a generalization of the
double projection method when a constraint operator is considered.
We discuss specific cases of quantum number projections for
the angular momentum, the momentum of the center of mass,
and the particle number.
In Sec. III, we apply the DGCM to BCS calculations and discuss the
effect of a fluctuation of mean particle numbers on the ground state of spherical
nuclei.
We then summarize the paper and discuss future perspectives
in Sec. IV.

\section{Dynamical GCM and generalization of the double projection method}
\subsection{A brief summary of DGCM}

In the generator coordinate method (GCM), one diagonalizes
a Hamiltonian $\hat{H}$ in the space spanned by
states $\{|\vec{q}\rangle\}$ which are parametrized by
generator coordinates $\vec{q}$.
Usually, many-body Slater determinants are used for the states $\{|\vec{q}\rangle\}$ with a real number $\vec{q}$.
Notice that the states  $\{|\vec{q}\rangle\}$ are not orthogonal to each other.
For simplicity, in the following, we consider only a single generator coordinate, $q$.
A many-body wave function is then expanded as
\begin{equation}
|\psi\rangle=\int dq\,f(q)|q\rangle.
\label{wfgcm}
\end{equation}
In this equation, the weight function $f(q)$ is determined by the variational principle,
which leads to the Hill--Wheeler equation \cite{Ring_Schuck},
\begin{equation}
  \int dq'\,(\langle q|\hat{H}|q'\rangle-E\langle q|q'\rangle)f(q')=0,
\end{equation}
where $E$ is an energy eigenvalue.
$\langle q|\hat{H}|q'\rangle$ and $\langle q|q'\rangle$ are referred to as
the Hamiltonian and the overlap kernels, respectively.
The GCM is often employed
to describe collective motions, and in this sense $q$ is called a
collective coordinate.

In principle,
if one could generate a collective coordinate properly,
the GCM could correctly describe a collective motion.
This is the case, e.g., for a system described by the Lipkin model \cite{Ring_Schuck}.
However, in general, it is an extremely difficult problem to
find properly a collective path, and one often
determines it in an empirical way.
There is no guarantee that the basis constructed in this way adequately takes
into account
the relevant dynamics of a collective motion which one wants to describe.

In order to overcome this problem,
Goeke and Reinhard have extended the GCM by introducing
the canonical momentum $p$ conjugate to the collective coordinate $q$
and defined the basis states which satisfy
\begin{equation}
 \label{conjugate}
  \langle q, p|
\overleftarrow{\partial}_{q} \overrightarrow{\partial}_{p}-\overleftarrow{\partial}_{p} \overrightarrow{\partial}_{q}| q, p\rangle=i.
\end{equation}
Here,
$\overleftarrow{\partial}_{q}$ and $\overrightarrow{\partial}_{q}$ act
on the left-hand and the right-hand sides of $q$, respectively, and similar for
$\overleftarrow{\partial}_{p}$ and $\overrightarrow{\partial}_{p}$.
Notice that we have set $\hbar=1$.
The path connecting $|q,p\rangle$ is called a dynamical path.

The condition  (\ref{conjugate}) can also be written in a form of the
commutation relation,
\begin{equation}
 \langle q, p|[\hat{Q}_0,\hat{P}_0]|q,p\rangle=i,
\end{equation}
where
$\hat{Q}_0$ and $\hat{P}_0$ are
generators of $q$ and $p$ defined as
\begin{eqnarray}
\hat{Q}_0|q,p\rangle&=&-\left(i\partial_p+\frac{\partial S}{\partial p}\right)|q,p\rangle, \\
\hat{P}_0|q,p\rangle&=&\left(i\partial_q+\frac{\partial S}{\partial q}\right)|q,p\rangle,
\end{eqnarray}
respectively, with an arbitrary smooth function, $S=S(q,p)$, of $q$ and $p$.
The function $S=S(q,p)$ originates from the freedom to choose any phase of
the state $|q,p\rangle$.
In Ref. \cite{GR80}, the phase was chosen
so that the expectation values of the two operators $\hat{Q}_0$ and $\hat{P}_0$
are zero.

After the dynamical path is somehow obtained,
one can expand a wave function using the states $|q,p\rangle$ as,
\begin{equation}
  |\psi\rangle= \iint dqdp\,f(q,p)|q,p\rangle.
\label{wfdgcm}
\end{equation}
This is called the dynamical GCM (DGCM)~\cite{GR78,RG78a,RG78b,RG78b,RG78c,GR80}.
The weight function
$f(q,p)$ is determined by solving the Hill--Wheeler equation,
as in the usual GCM.

In general, not all the states specified by the two parameters $q$ and $p$
contribute to a collective motion.
For example, if one could find a function $\Gamma(q,p;q')$ which satisfies
\begin{equation}
  |q,p\rangle=\int dq'\,\Gamma(q,p;q')|q',0\rangle
\end{equation}
for an arbitral pair of $(q,p)$, the wave function in the DGCM, Eq. (\ref{wfdgcm}),
is reduced to the wave function in the GCM, Eq. (\ref{wfgcm}), with
\begin{equation}
f_{\rm GCM}(q')=\iint dpdq\,f_{\rm DGCM}(q,p)\Gamma(q,p;q').
\end{equation}
Here,
$f_{\rm GCM}(q)$ and $f_{\rm DGCM}(q,p)$ are the weight functions in Eqs.
(\ref{wfgcm}) and (\ref{wfdgcm}), respectively.
In this case, there is no need to consider the DGCM and the usual GCM is
sufficient. This condition is called the global redundancy\cite{GR80}.
Goeke and Reinhard further showed that
there are certain cases in which
a one-parameter GCM along a path in the $(q,p)$ space
suffices even without the global redundancy \cite{GR80}.
However, in general,
the configuration along such a relevant path is considerably complicated. In that
situation, one can instead apply the DGCM in a straightforward manner.
Nevertheless,
it is not numerically easy to construct the configurations
along a dynamical path,
partly because
the number of collective variables is doubled in the DGCM.
For this reason, the DGCM has not yet been applied to any concrete numerical
problems.

\subsection{Dynamical path from a constrained Hartree--Fock method}

Another potential problem of the DGCM is that it is not obvious how to practically
find a dynamical path. In this regard, we show below that there is a reasonable way to
construct a dynamical path when a collective coordinate is generated
by the constrained Hartree--Fock(--Bogoliubov) method.

Suppose that $|q\rangle$ is a many-body state which satisfies
\begin{equation}
\langle q|\hat{Q}_0|q\rangle=q
\end{equation}
with a Hermitian operator $\hat{Q}_0$.
We then define a state
\begin{equation}
  |q,p\rangle\equiv e^{i\hat{Q}_0p}|q\rangle.
\label{qpdef}
\end{equation}
It is obvious that this state satisfies
\begin{equation}
  \langle q,p|\hat{Q}_0|q,p\rangle=q.
\end{equation}
Differentiating both sides of this equation by $q$, we then
obtain
\begin{equation}
\langle q,p|\overleftarrow{\partial}_{q}\hat{Q}_0+\hat{Q}_0\overrightarrow{\partial}_{q}|q,p\rangle=1.
\end{equation}
Noticing
\begin{equation}
i\hat{Q}_0  |q,p\rangle=
\overrightarrow{\partial}_{p}|q,p\rangle,
\end{equation}
which follows from the definition of the state $|q,p\rangle$, (\ref{qpdef}),
we find
\begin{equation}
  \langle q, p|\overleftarrow{\partial}_{q} \overrightarrow{\partial}_{p}-\overleftarrow{\partial}_{p} \overrightarrow{\partial}_{q}| q, p\rangle=i.
\end{equation}
This is nothing more than the conjugate condition, Eq. (\ref{conjugate}).
That is, when one uses the constrained Hartree--Fock(--Bogoliubov) method
to generate a collective coordinate
with a Hermitian operator, one can always
construct a desired dynamical path.

If one employs the states $|q,p\rangle$ so obtained in the DGCM,
a many-body wave function is expressed as,
\begin{equation}
  \label{DGCM}
  |\psi\rangle=\iint dqdp\,f(q,p)e^{i\hat{Q}_0p}|q\rangle.
\end{equation}
We rewrite this equation using the Fourier transform of $f(q,p)$,
\begin{equation}
f(q,p)=\int dq'\,\tilde{f}(q,q')e^{-iq'p}.
\end{equation}
This leads to
\begin{equation}
|\psi\rangle=\iint dqdq'\,\tilde{f}(q,q')\left \{\int dp\,e^{i(\hat{Q}_0-q')p}\right\}|q\rangle.
\label{wfdgcm2}
\end{equation}
Notice that, apart from the normalization coefficient,
\begin{equation}
\hat{P}_{q'}^{(\hat{Q}_0)}\equiv\int dp\,e^{i(\hat{Q}_0-q')p}
\end{equation}
is the projection operator which projects a state onto an eigenfunction of the
operator $\hat{Q}_0$ with an eigenvalue of $q'$.
Thus, the DGCM state can be expressed as.
\begin{equation}
  \label{DP}
  |\psi\rangle=\iint dqdq'\,\tilde{f}(q,q')
\hat{P}_{q'}^{(\hat{Q}_0)}|q\rangle.
\end{equation}
This implies that the DGCM is equivalent to
the GCM supplemented by a projection method.

\subsection{Translational motion}

Let us apply
the formula derived in the previous subsection
to the center of mass motion and compare with the
double projection method.
To this end, we first generate the state $|\vec{p}\rangle$ which satisfies
\begin{equation}
  \langle \bm{p}|\hat{\bm{P}}|\bm{p}\rangle = \bm{p},
\end{equation}
where $\hat{\bm{P}}=(\hat{P}_x,\hat{P}_y,\hat{P}_z)$ is the operators for the
center of mass motion of a whole system.
Since
$\hat{P}_x$, $\hat{P}_y$, and $\hat{P}_z$ commute with each other,
the wave function in the DGCM, Eq. (\ref{wfdgcm2}), reads,
\begin{equation}
\label{mom_DGCM}
|\psi\rangle=\iint d\vec{p}'d\vec{p}''\,\tilde{f}
(\bm{p}',\bm{p}'')\left \{\int d\vec{q}\,e^{-i(\hat{\bm{P}}-\bm{p}'')\cdot \bm{q}}\right\}|\bm{p}'\rangle.
\end{equation}
Since the operators $\hat{\bm{P}}$ commutes with the Hamiltonian,
one would be interested only in the eigenstates of
$\hat{\bm{P}}$.
Acting the projection operator for the operator $\hat{\vec{P}}$
onto Eq. (\ref{mom_DGCM}), one then
obtains
\begin{equation}
  \label{mom_DP}
  |\psi\rangle_{\bm{p}}=\iint d\vec{q}d\vec{p}'\,\tilde{f}(\bm{p}',\bm{p})
e^{-i(\hat{\bm{P}}-\bm{p})\cdot \bm{q}}|\bm{p}'\rangle.
\end{equation}
This coincides with
Eq. (2.6) in Ref. \cite{PT62}.
In this way,
the ansatz of the double projection method can be directly derived from the DGCM,
when
the operators to generate a generator coordinate commutes with a Hamiltonian.
In this sense,
the DGCM can be regarded as an extension of the double projection method
of Peierls and Thouless \cite{PT62}.

\subsection{Rotational motion}

Let us next consider a rotational motion, thus, the angular momentum.
For the sake of simplicity, we consider only a
rotation around the $x$-axis.
Using the operator $\hat{J}_x$,
we first generate a gerator coordinate imposing a condition of
\begin{equation}
  \label{angconst}
  \langle m|\hat{J}_x|m\rangle = m.
\end{equation}
Following the same procedure as in the center of mass motion,
one can write the DGCM anzatz for the eigenstates of
$\hat{J}_x$ as
\begin{equation}
  |\psi\rangle_{m}=\iint d\theta dm'\,f(m',m) e^{-i(\hat{J}_x-m)\theta}|m'\rangle.
\end{equation}
This is also consistent with Eq. (3.11) in Ref. \cite{PT62}.

Unfortunately, it is not straightforward to extend this discussion to
a general rotation, since
the angular momentum operators $\hat{J}_x$, $\hat{J}_y$, and $\hat{J}_z$
do not commute with each other, unlike the linear momentum operators, $\hat{\vec{P}}$.
One possible prescription is to construct a DGCM wave function using
$\hat{J}_x$ and $\hat{\bm{J}}^2$.
Wave functions similar to this have been considered in Refs.
\cite{BR15,EB16,ST15,Shimada2016,Ushitani2019}, in which
the following anzatz was employed:
\begin{equation}
\label{angGCM}
|LM\rangle=\sum_{K,m'}f_{Km'}\hat{P}_{MK}^L|m'\rangle.
\end{equation}
Here,
$m'$ is the gerator coordinate defined by (\ref{angconst}) and
$\hat{P}^L_{MK}$ is the angular momentum projection operator.
We have dropped other parameters than the angular momentum, such as
deformation, from the notation in Eq. (\ref{angGCM}).
Notice that
the idea of DGCM is applied in this equation only to the angular momentum
component in the direction of the quantization axis, while
the effect of the fluctuation of the total angular momentum is not considered.
It might be an interesting future work to extend this prescription by
introducing a generator coordinate associated with the total angular momentum
in addition to that in Eq. (\ref{angconst}).

\subsection{Particle number}

We next consider the particle number projection.
Using the particle number operator $\hat{N}$, we
first generate a generator coordinate according to
\begin{equation}
  \langle N|\hat{N}|N\rangle=N.
\end{equation}
Here, the state $|N\rangle$ represents either a BCS state or a
Hartree--Fock--Bogoliubov state, in which several particle number components
are mixed.
If we construct the eigenstate of $\hat{N}$,
one can write the DGCM wave function as
\begin{equation}
|\psi\rangle_{N_0}=\iint dN d\phi \,f(N,N_0) e^{i(\hat{N}-N_0)\phi}|N\rangle.
\end{equation}
Here,
$\phi$ is the gauge angle, which is a quantity conjugate to the particle number.
If one considers a non-relativistic case,
the particle number operator is semi-positive definite.
In this case,
the integral range of $N$ is from 0 to $\infty$.
On the other hand,
for $\phi$, the range of the integral is from 0 to 2$\pi$.
Using the particle number projection operator defined by
\begin{equation}
\hat{P}^{N_0}=\int_0^{2\pi}\frac{d\phi}{2\pi}\,e^{i(\hat{N}-N_0)\phi},
\label{NumberProjection}
\end{equation}
one thus has
\begin{equation}
  \label{NP-GCM}
|\psi\rangle_{N_0}=\int_0^{\infty}dN\,f_{N_0}(N)\hat{P}^{N_0}|N\rangle,
\end{equation}
except for a normalization constant.
This can also be interpreted as the double
projection method for the particle number fluctuation.
The Hill--Wheeler equation for $f_{N_0}(N)$ reads
\begin{equation}
  \int_0^{\infty} dN'\,\left(\langle N|\hat{H}\hat{P}^{N_0}|N'\rangle-E\langle N|\hat{P}^{N_0}|N'\rangle\right)f_{N_0}(N')=0.
\end{equation}

Notice that, in the case of the particle number,
the usual GCM corresponds to the
variation before projection (VBP) method,
\begin{equation}
  |\psi\rangle_N\propto\int_0^{2\pi} \frac{d\phi}{2\pi}\,e^{i(\hat{N}-N)\phi}|N\rangle.
\end{equation}
In addition to the gauge angle $\phi$, if one treats $N$ as a generator
coordinate, one obtains the DGCM wave function, (\ref{NP-GCM}).
While the VBP takes into account only the fluctuation of the gauge angle,
the DGCM incorporates the effect of the fluctuation of a mean particle number
in mean-field wave functions.

Even though the DGCM, or the double projection method, has
not been applied to the particle number fluctuation,
GCM calculations based on a similar idea
have been carried out treating the pairing fluctuation as a
generator coordinate \cite{VR11,FYNB,Vaquero2013}. See also Ref. \cite{Giuliani2014}.
There, the generating functions are constructed as
\begin{equation}
  |\langle\delta| (\Delta\hat{N})^2 |\delta\rangle|^{1/2} = \delta,
\end{equation}
using the operator $\Delta\hat{N}=\hat{N}-\langle\hat{N}\rangle$.
Following the idea of GCM, these wave functions are linearly superposed as
\begin{equation}
  |\psi\rangle_{N_0}=\int d\delta f(\delta)\hat{P}^{N_0}|\delta\rangle.
\end{equation}
It has been shown that such treatment of the pair fluctuation improves the
description of the structure of $^{54}$Cr \cite{VR11} and also significantly
affects nuclear matrix elements of double beta decays \cite{Vaquero2013}.
This method indeed takes into account the effect of pairing fluctuation,
but it is not clear whether it fully takes it into account in a sense of the
DGCM illustrated in this subsection.

\section{Numerical calculations for DGCM for particle number}

\subsection{Numerical details}

In this section, we apply the DGCM to actual nuclei
and numerically investigate the effect of the particle number fluctuation in
a BCS wave function using Eq.~(\ref{NP-GCM}).
To this end, we focus for simplicity only on the neuron number.
We thus choose the singly closed
$^{16,18}$O, $^{40,42}$Ca, and $^{56,58,64}$Ni nuclei and assume that the protons
are in the normal fluid phase.
Considering the systematic calculation~\cite{erl12},
we also assume that these nuclei have a spherical symmetry.
We employ the
\mbox{SI\hspace{-.1em}I\hspace{-.1em}I}
Skyrme energy functional~\cite{BF75}.

We prepare a set of many-body wave functions $|N\rangle$ which
have the average particle number of $N$. Notice that $N$ may be different from
the actual neutron number $N_0$ for each nucleus, and that $N$ may not necessarily
be an integer number.
For this purpose, we
employ the BCS approximation for the
pairing correlation among neutrons, while we ignore the neutron--proton
pairing.
We solve the Skyrme--Hartree--Fock equation in a box with 15 fm with a mesh spacing of 0.05 fm,
and the continuum states are then discretized.
To calculate the pairing energy,
we employ the
pairing energy functional given by
\begin{equation}
E_{\mathrm{pair}}[\rho,\tilde{\rho}]=\frac{V_n}{4}\int d\vec{r}
\left(1-\frac{\rho(\bm{r})}{\rho_0}\right)|\tilde{\rho}_n(\bm{r})|^2,
\end{equation}
where $\rho(\vec{r})$ and
$\tilde{\rho}_n(\bm{r})$ are the total particle density and the neutron pair density,
respectively, with $\rho_0$ being 0.16 fm$^{-3}$.
The pairing energy is calculated with an
energy cut-off at 15 MeV above the Fermi energy.

For $^{18}$O, $^{42}$Ca, and $^{58,64}$Ni, we determine the value of $V_n$
so that the average pairing gap,
\begin{equation}
\bar{\Delta}=\frac{\frac{V_n}{2}\int d^3r\left(1-\frac{\rho(\bm{r})}{\rho_0}\right)|\tilde{\rho}_n(\bm{r})|^2}{\int d^3r|\tilde{\rho}_n(\bm{r})|^2},
\end{equation}
coincides with the empirical value, $12/\sqrt{A}$ MeV, where $A$ is the mass number of
a nucleus.
For the doubly magic nuclei, $^{16}$O and $^{40}$Ca, we somewhat reduce the value of
$V_n$ so that the pairing gap becomes zero in the BCS approximation. For the
$^{56}$Ni nucleus, this problem does not appear and we use the same value
of $V_n$ as that for $^{58}$Ni.
The parameters are listed in Tab.~\ref{tab:pairing_data}.

\begin{table*}[btp]
\caption{The strengths of the pairing interaction, $V_n$, employed in
the present calculations. These are given in units of MeV fm$^{-3}$.}
\label{tab:pairing_data}
\centering
\small
\begin{tabular}{cccccccc}
\hline\hline
nucleus & $^{16}$O & $^{18}$O & $^{40}$Ca & $^{42}$Ca &
   $^{56}$Ni  &  $^{58}$Ni & $^{64}$Ni \\
   \hline
$V_n$ & $-$800.00 & $-$901.98 & $-$700.00 & $-$775.23 & $-$897.80 & $-$897.80 & $-$707.81 \\
    \hline\hline
  \end{tabular}
\end{table*}

We mainly show below the results with $\Delta N=0.2$ around $N_0$ in the range
of $N_0-2\leq N \leq N_0+2$.
For simplicity, for each nucleus
we ignore the non-orthogonality of single-particle wave functions
for different values of $N$ and
assume $\langle \varphi_i(N)|\varphi_j(N')\rangle=\delta_{i,j}$, where $\varphi_i(N)$ is
the $i$-th single-particle wave function for a system with the average neutron
number of $N$. We have confirmed that the deviation from this condition is negligibly
small in the range of $N$ considered in this paper.
We thus use the single-particle wave functions for $N=N_0$ for each nucleus.

\begin{figure} [bthp]
  \includegraphics[scale=0.74,clip]{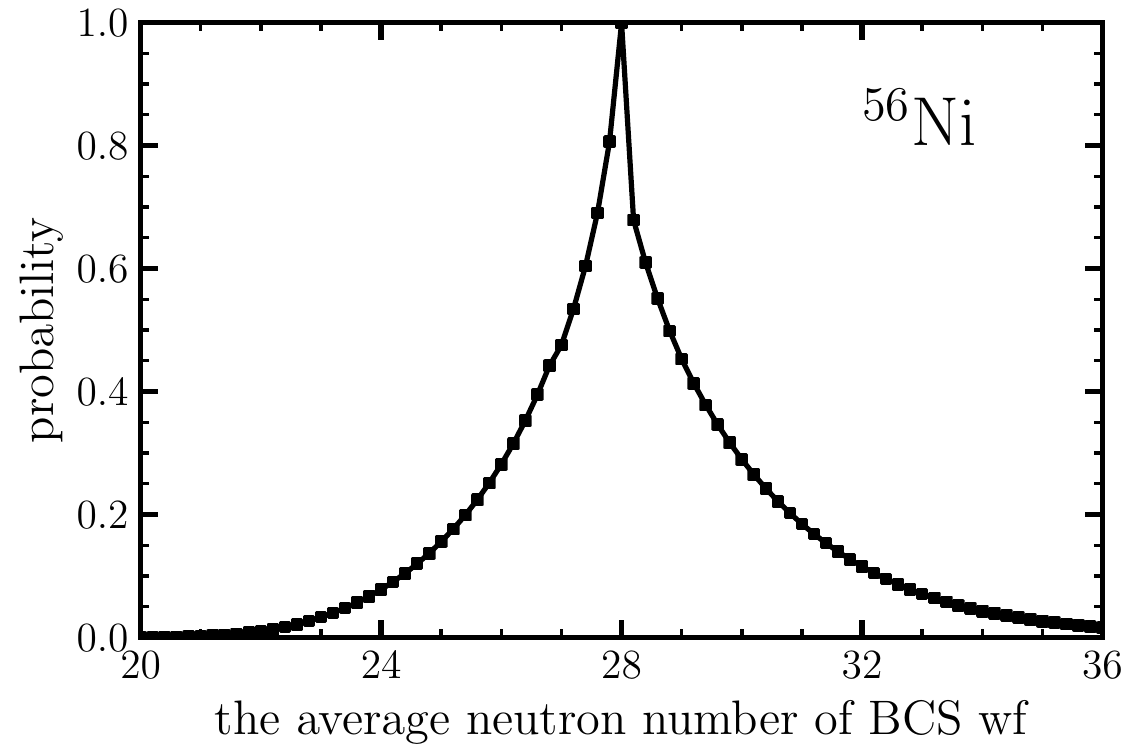}
\caption{The probability to find the $N_0=28$ component in the BCS wave function
$|N\rangle$ for $^{56}$Ni which has the average neutron number of $N$.
  }
  \label{fig:fluct}
\end{figure}

Figure~\ref{fig:fluct} shows
the probability of the component of $N_0$ in each BCS wave function $|N\rangle$
as a function of $N$ for the $^{56}$Ni nucleus ($N_0=28$).
This is computed as
$P^{N_0}_N=|\langle N|\hat{P}^{N_0}|N\rangle|^2$,
with the particle number projection operator, (\ref{NumberProjection}).
One can see that the probability has a large value in the range considered in
this paper, $N_0-2\leq N\leq N_0+2$.
The BCS states with larger values of $N$ have a smaller overlap with the state with
$N_0$, and inclusion of such states in the DGCM
may cause a serious numerical problem.

We then apply the particle number projection, (\ref{NumberProjection}),
to the wave functions $|N\rangle$ and superpose them according to Eq.~(\ref{NP-GCM}).
For this purpose,
we discretize the gauge angle $\phi$ with
$\Delta\phi=2\pi/80$ for the integral with respect to $\phi$.
We use the mixed density prescription
to calculate the Hamiltonian and the overlap kernels \cite{DS07,BD90}.
Since we use the same single-particle wave functions for each $N$,
the mixed density and the mixed pair density are simply given by
\begin{eqnarray}
 \rho^\phi_{NN'}(\bm{r})&=&\sum_{i}\frac{v_i^Nv_i^{N'}e^{2i\phi}}
{u_i^Nu_i^{N'}+v_i^Nv_i^{N'}e^{2i\phi}}|\varphi_i(\bm{r})|^2, \\
 \tilde{\rho}^\phi_{NN'}(\bm{r})&=&\sum_{i}\frac{u_i^Nv_i^{N'}e^{2i\phi}}
{u_i^Nu_i^{N'}+v_i^Nv_i^{N'}e^{2i\phi}}|\varphi_i(\bm{r})|^2,
\end{eqnarray}
respectively.
Here,
$u_i^{N}$ and $v_i^{N}$ are the $uv$-factors for the single-particle state $i$
in the BCS wave function with
the average neutron number $N$.
Other local mixed densities are given in a similar way.

In our calculations, we superpose many similar states.
The problem of overcompliteness may then arise~\cite{Ring_Schuck}
due to the linear dependence of the bases.
To avoid this problem,
in numerical calculations shown below,
we remove
the eigenstates of the overlap kernel whose eigenvalue is smaller than
$\lambda_{\mathrm{cut}}=10^{-5}$ (see Fig. 2 below for the dependence of
the result on the choice of $\lambda_{\mathrm{cut}}$).
In the actual calculations,
with this remedy for the overcompliteness,
we use the subroutine DSYEV of the LAPACK package~\cite{LAPACK}
to diagonalize the discretized Hill--Wheeler equation
as both the Hamiltonian and the overlap kernels are real symmetric
matrices in the present calculation.

\subsection{Results}

\begin{figure} [btp]
\includegraphics[scale=0.74,clip]{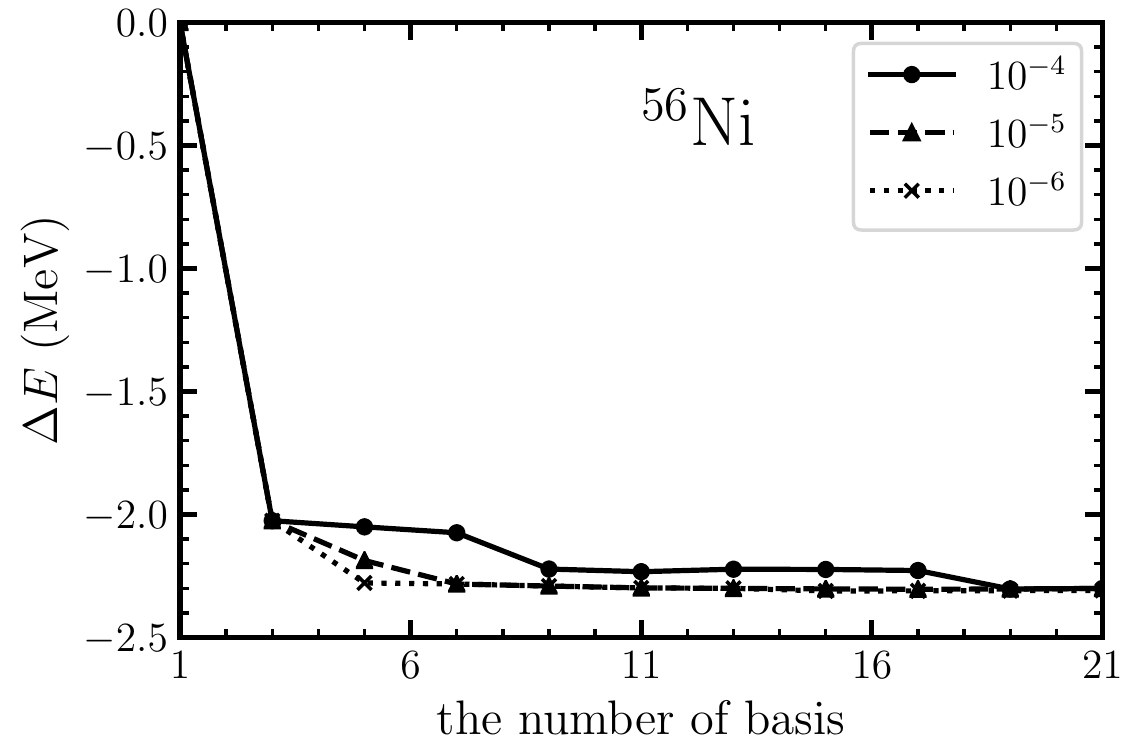}
\caption{The energy gain in the $^{56}$Ni nucleus as a function of the number
of basis states $N_{\rm DGCM}$ in the DGCM method.
It is plotted with respect to the energy of $N_{\rm DGCM}=1$, which is equivalent
to the variation before projection method.
The solid, the dashed, and the dotted lines denote the results
with $\lambda_{\rm cut}=10^{-4},~10^{-5}$, and 10$^{-6}$ for
the cut-off of the eigenvalues of the overlap kernel, respectively.
}
\label{fig:cut-off}
\end{figure}

Figure \ref{fig:cut-off} shows the total energy gain
$\Delta E$ for $^{56}$Ni
due to the superposition of various $|N\rangle$ states in Eq.~(\ref{NP-GCM}).
Here, the enegy gain is defined as
$\Delta E=E(N_{\mathrm{DGCM}})-E(N_{\mathrm{DGCM}}=1)$, where
$E(N_{\mathrm{DGCM}})$ is the total energy of the system
when the number of basis is $N_{\mathrm{DGCM}}$.
This quantity is plotted as a function of the number of basis ($|N\rangle$),
$N_{\rm DGCM}$, where $N_{\rm DGCM}=1$ corresponds to the usual variation before particle
number projection (VBP).
To draw the figure, we increase the number of basis
by adding two basis states symmetrically around $N_0$, that is,
$N_0,~N_0\pm\Delta N,~N_0\pm2\Delta N \dots$.
The solid, the dashed, and the dotted lines denote the results
with $\lambda_{\rm cut}=10^{-4},~10^{-5}$, and 10$^{-6}$ for
the cut-off of the eigenvalues of the overlap kernel, respectively.
One can see that the results are almost converged at $\lambda_{\rm cut}=10^{-5}$.
We thus use this value
in all the calculations shown below unless otherwise mentioned.
The figure also shows that the energy gain quickly converges as a function
of $N_{\rm DGCM}$. In particular, the energy is significantly decreased even
with a mixture of three basis states only, $N_{\rm DGCM}=3$.
We have repeated the same calculation with $\Delta N=0.1$ and
have found that the converged energy remains almost the same as that
with $\Delta N=0.2$, with
a similar convergence feature to each other.

\begin{table}[tbp]
\caption{The ground-state energy of each nucleus calculated with different methods.
The energies are given in units of MeV.
For the DGCM method, the number in the parenthesis denotes the number of
basis states, $N_{\mathrm{GCM}}$, for which DGCM(1) is equivalent to VBP. }
  \label{tab:energy}
  \centering
  \begin{tabular}{ccccc}
    \hline
    \hline
    \qquad\qquad  & BCS & DGCM(1) & DGCM(3) & DGCM(21) \\
    \hline
    $^{16}$O  & $-$128.01 & $-$128.01 & $-$128.41 & $-$129.29 \\
    $^{18}$O  & $-$144.91 & $-$147.50 & $-$147.82 & $-$148.03 \\
    $^{40}$Ca & $-$341.30 & $-$341.30 & $-$342.63 & $-$342.79 \\
    $^{42}$Ca & $-$363.83 & $-$365.55 & $-$365.87 & $-$365.98 \\
    $^{56}$Ni & $-$482.74 & $-$482.74 & $-$484.76 & $-$485.04 \\
    $^{58}$Ni & $-$504.40 & $-$506.31 & $-$507.37 & $-$507.79 \\
    $^{64}$Ni & $-$557.87 & $-$559.44 & $-$559.66 & $-$559.84 \\
    \hline
    \hline
  \end{tabular}
\end{table}

\begin{figure} [tbp]
  \includegraphics[scale=0.74,clip]{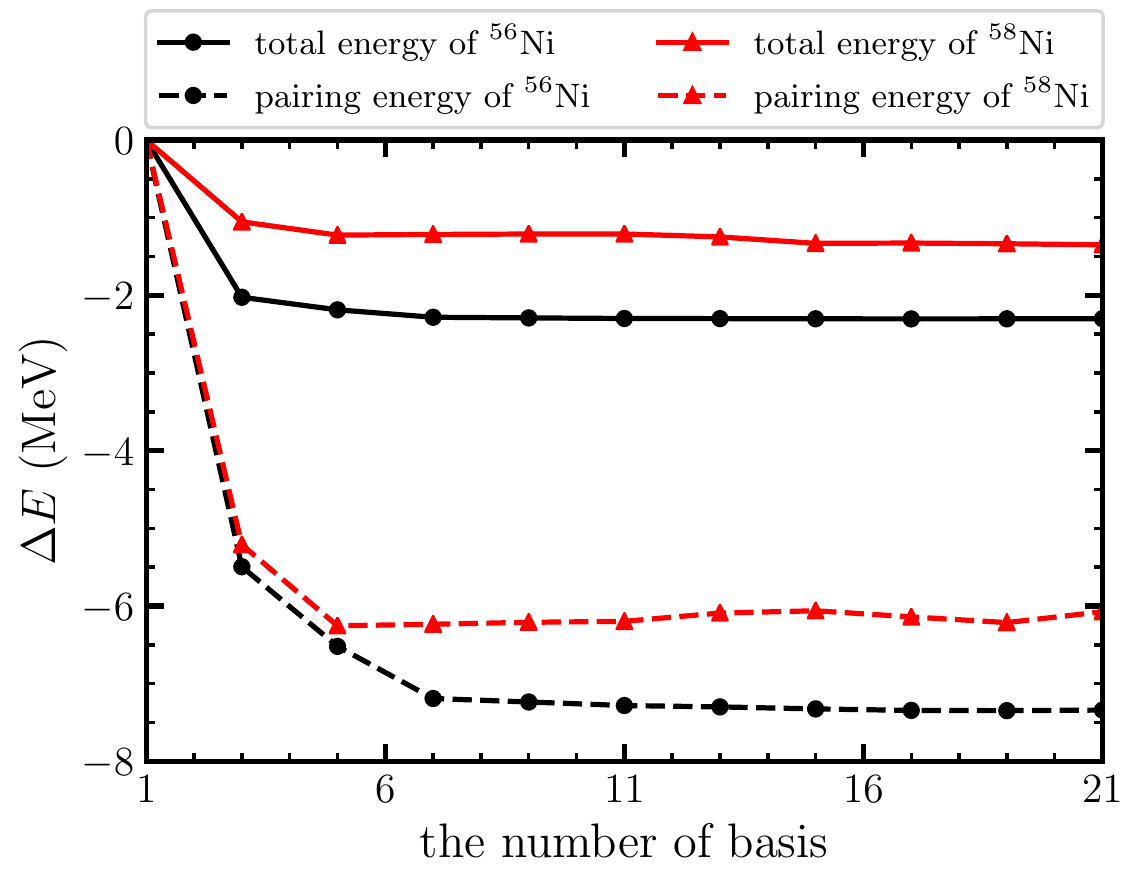}
  \caption{Similar to Fig. \ref{fig:cut-off}, but for a comparison between the
total energy (the solid lines) and the pairing energy (the dashed lines).
The filled circles and the filled triangles denote the results for the
$^{56}$Ni and the $^{58}$Ni nuclei, respectively. }
  \label{fig:pairing}
\end{figure}

\begin{figure} [tbhp]
  \includegraphics[scale=0.74,clip]{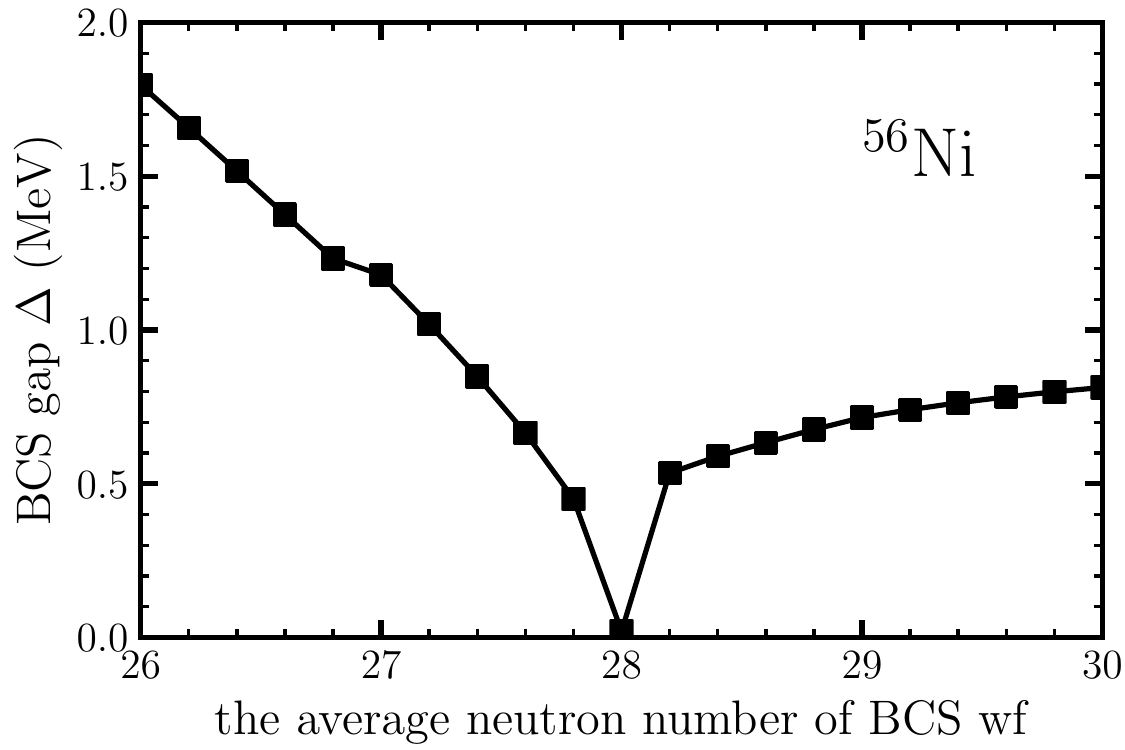}
\caption{The pairing gap $\Delta$ of $^{56}$Ni in the BCS approximation
for the basis states for the DGCM calculation.
}
\label{fig:gap}
\end{figure}

Table~\ref{tab:energy} summarizes the results for the
$^{16,18}$O, $^{40,42}$Ca, and $^{56,58,64}$Ni nuclei
\footnote{For $^{16}$O, the problem of overcompliteness is found to be severe,
and we chose $\lambda_{\mathrm{cut}}=8.0\times10^{-2}$, which is
determined from the eigenvalue distribution of the overlap kernel.}.
One can see that a large energy gain is obtained for all of these cases, as in
$^{56}$Ni shown in Fig. \ref{fig:cut-off}.
As we have discussed in Sec. II-D, this can be interpreted as
a consequence of the fluctuation of a mean
particle number in mean-field wave functions.
It is noteworthy that the energy gain is particularly large for
the neutron magic nuclei,
$^{16}$O, $^{40}$Ca, and $^{56}$Ni.
To clarify the reason for this, we show in Fig.~\ref{fig:pairing}
the energy gain (the solid lines) and the contribution of the pairing energy (the
dashed lines)
as a function of $N_{\rm DGCM}$.
The lines with the filled circles denote the results for $^{56}$Ni, while the lines with the
filled triangles are for $^{58}$Ni.
One can clearly see that
the total energy decreases with the
development of the pairing energy.
It is interesting to notice that the pairing contribution is larger in
the neutron magic nucleus $^{56}$Ni as compared to that in $^{58}$Ni.
This is due to the fact that, for $^{58}$Ni, the effect of the pairing
correlation is already
taken into account to some extent
in the calculation with $N_{\rm DGCM}=1$, while for $^{56}$Ni
the energy with VBP does not change from that in the BCS approximation due to the
absence of the pairing gap.
To illustrate this,
Fig. \ref{fig:gap} shows the BCS pairing gap
for the basis states $|N\rangle$ for $^{56}$Ni used in this study.
While the pairing gap is zero for $N=28$, the gap is finite for other basis states.
Therefore, this nucleus can take an advantage of finite pairing gaps by mixing
configurations with $\langle \hat{N}\rangle\neq 28$, which significantly lowers
the total energy.
In this sense, the DGCM for the particle number is somewhat similar to
the GCM calculations where a pairing fluctuation is treated as a
generator coordinate.
It is also noted that
the energy gain due to DGCM is small for nuclei where the pairing
correlation is well developed, such as $^{64}$Ni shown in Tab.~\ref{tab:energy}.

In the mean-field calculations, the
variation after projection (VAP) method is more consistent
than the VBP method~\cite{Ring_Schuck,Sheikh2000,Sheikh2002,Stoitsov2007,Duguet2009,Hupin2012}.
However, the VAP is much more cumbersome and is often numerically more involved
as compared to the VBP.
One may resort to the Lipkin--Nogami method (LN) \cite{L60,N64} as an approximation
of the VAP, but it has been know that the LN method does not work well
for nuclei closed to shell closures  \cite{ZS92,DN93,HB00,Hagino2002}.
The method proposed in this paper is much simpler than the VAP, yet
a similar amount of the energy gain
can be obtained with a lower computation cost. In particular, it is
a numerical advantage of our method that the total energy is significantly
lowered already with $N_{\rm DGCM}=3$. Moreover, our method works well
not only for open shell nuclei but also for nuclei close to a shell closure.
We thus argue that our method can be a good alternative to
the VAP and the Lipkin--Nogami methods.

\section{Summary and future perspectives}

We have discussed an extension of the generator coordinate method (GCM) by treating
both a collective coordinate and its conjugate momentum as generator coordinates.
To this end, we have investigated the idea of the dynamical GCM (DGCM).
We have first shown that a dynamical path relevant to the DGCM can be
constructed whenever
a collective coordinate is generated by the constrained mean-field method
with a Hermitian operator.
The DGCM can thus be applied once the operator relevant to a collective
motion is identified.
In such cases, the DGCM
can be formulated in a form of a generalized double projection method.

We have applied the DGCM to the particle number projection as an example.
Here,
we have superposed many BCS states which have different mean particle numbers,
after performing the particle number projection.
In this way,
we have incorporated not only the fluctuation of the gauge angle but also
the fluctuation of the mean particle number in the BCS wave functions.
As a result, the ground state energy was found to be significantly lowered
compared to the BCS+VBP method, which can be regarded as the usual GCM
for the gauge angle.
In particular, for magic nuclei,
no energy gain is obtained in the VBP method due to a vanishing of pairing
correlation in the mean field approximation, while this method yields
a significant energy gain by mixing configurations with a non-zero
pairing gap.
This consideration may be important for
the Mottelson--Valatin effect~\cite{MV60} in nuclear superconductivity
at high angular momenta and/or at high temperatures.
There, the fluctuation beyond the mean field approximation
may play an important role~\cite{ER85,SB90}, and
it may be interesting to recast this problem from the
viewpoint of DGCM.

We have also found that the convergence of the DGCM calculations
is fast with respect to the number
of superposed states, and a considerable energy gain
is obtained by mixing only three configurations.
The method is much simpler than the VAP, and
thus a numerical calculation is much easier. Moreover, the method works well
for both open shell nuclei and magic nuclei, unlike the Lipkin--Nogami method
which does not work for nuclei at shell closures.
We thus advocate this method as a good alternative to the VAP method.

In this paper, for simplicity we have carried out all the calculations
assuming spherical symmetry. It would be an interesting future work
to extend this by removing the restriction of nuclear shape. This would
be important particularly for soft nuclei, for which the nuclear shape may
change significantly as a function of the average particle number.
In such cases, it is an advantage of our method that correlations associated with
the shape degree of freedom can be largely incorporated by mixing a few
configurations with different average particle numbers.

In addition, even though we have focused in this paper only on the neutron number,
it is straightforward to include the fluctuation of the average proton number in
BCS states as well.
The resultant DGCM wave function can easily incorporate the effect of a
proton--neutron pairing, because
it is no longer a form of the direct product of proton and neutron states.
This may be a good advantage to explaining the Wigner energy \cite{W37, IW95}, for which the fluctuation of
a pair field may play an important role \cite{Satula1997,Satula1997-2}.

Our calculations presented in this paper indicates that a better description
can be achieved with the DGCM by incorporating the conjugate momentum of a collective
coordinate. This would mean that GCM calculations reported in the literature
may need to be reexamined from the view point of the DGCM.
For example, for a GCM calculation with
the quadrupole deformation operator $\hat{Q}_{20}$,
the DGCM wave function can be constructed according to
Eqs. (\ref{DGCM}) or (\ref{DP}).
This would require a development of a computation method
to evaluate the operator in a form of $\exp(iq\hat{Q}_{20})$ or
to carry out the quantum number projection for the operator
$\hat{Q}_{20}$.
Since the multipole operators are mutually commutative,
the DGCM wave function can be easily extended to a multi-dimensional deformation
plane as well.
It would be an interesting future problem to
develop a microscopic fission theory \cite{Bender2020} based on this idea.

\section*{Acknowledgments}
The authors thank N.~Hinohara for useful discussions.
This work was supported by JSPS KAKENHI
(Grant Nos. JP19K03824, JP19K03861, and JP19K03872).

\bibliographystyle{apsrev4-2}
\bibliography{paper}

\begin{thebibliography}{67}%
\makeatletter
\providecommand \@ifxundefined [1]{%
 \@ifx{#1\undefined}
}%
\providecommand \@ifnum [1]{%
 \ifnum #1\expandafter \@firstoftwo
 \else \expandafter \@secondoftwo
 \fi
}%
\providecommand \@ifx [1]{%
 \ifx #1\expandafter \@firstoftwo
 \else \expandafter \@secondoftwo
 \fi
}%
\providecommand \natexlab [1]{#1}%
\providecommand \enquote  [1]{``#1''}%
\providecommand \bibnamefont  [1]{#1}%
\providecommand \bibfnamefont [1]{#1}%
\providecommand \citenamefont [1]{#1}%
\providecommand \href@noop [0]{\@secondoftwo}%
\providecommand \href [0]{\begingroup \@sanitize@url \@href}%
\providecommand \@href[1]{\@@startlink{#1}\@@href}%
\providecommand \@@href[1]{\endgroup#1\@@endlink}%
\providecommand \@sanitize@url [0]{\catcode `\\12\catcode `\$12\catcode
  `\&12\catcode `\#12\catcode `\^12\catcode `\_12\catcode `\%12\relax}%
\providecommand \@@startlink[1]{}%
\providecommand \@@endlink[0]{}%
\providecommand \url  [0]{\begingroup\@sanitize@url \@url }%
\providecommand \@url [1]{\endgroup\@href {#1}{\urlprefix }}%
\providecommand \urlprefix  [0]{URL }%
\providecommand \Eprint [0]{\href }%
\providecommand \doibase [0]{https://doi.org/}%
\providecommand \selectlanguage [0]{\@gobble}%
\providecommand \bibinfo  [0]{\@secondoftwo}%
\providecommand \bibfield  [0]{\@secondoftwo}%
\providecommand \translation [1]{[#1]}%
\providecommand \BibitemOpen [0]{}%
\providecommand \bibitemStop [0]{}%
\providecommand \bibitemNoStop [0]{.\EOS\space}%
\providecommand \EOS [0]{\spacefactor3000\relax}%
\providecommand \BibitemShut  [1]{\csname bibitem#1\endcsname}%
\let\auto@bib@innerbib\@empty
%</preamble>
\bibitem [{\citenamefont {Bender}\ \emph
  {et~al.}(2003{\natexlab{a}})\citenamefont {Bender}, \citenamefont {Heenen},\
  and\ \citenamefont {Reinhard}}]{Bender2003}%
  \BibitemOpen
  \bibfield  {author} {\bibinfo {author} {\bibfnamefont {M.}~\bibnamefont
  {Bender}}, \bibinfo {author} {\bibfnamefont {P.-H.}\ \bibnamefont {Heenen}},\
  and\ \bibinfo {author} {\bibfnamefont {P.-G.}\ \bibnamefont {Reinhard}},\
  }\href {https://doi.org/10.1103/RevModPhys.75.121} {\bibfield  {journal}
  {\bibinfo  {journal} {Rev. Mod. Phys.}\ }\textbf {\bibinfo {volume} {75}},\
  \bibinfo {pages} {121} (\bibinfo {year} {2003}{\natexlab{a}})}\BibitemShut
  {NoStop}%
\bibitem [{\citenamefont {Nikšić}\ \emph {et~al.}(2011)\citenamefont
  {Nikšić}, \citenamefont {Vretenar},\ and\ \citenamefont {Ring}}]{NV11}%
  \BibitemOpen
  \bibfield  {author} {\bibinfo {author} {\bibfnamefont {T.}~\bibnamefont
  {Nikšić}}, \bibinfo {author} {\bibfnamefont {D.}~\bibnamefont {Vretenar}},\
  and\ \bibinfo {author} {\bibfnamefont {P.}~\bibnamefont {Ring}},\ }\href
  {https://doi.org/https://doi.org/10.1016/j.ppnp.2011.01.055} {\bibfield
  {journal} {\bibinfo  {journal} {Prog. Part. Nucl. Phys.}\ }\textbf {\bibinfo
  {volume} {66}},\ \bibinfo {pages} {519 } (\bibinfo {year}
  {2011})}\BibitemShut {NoStop}%
\bibitem [{\citenamefont {Egido}(2016)}]{E16}%
  \BibitemOpen
  \bibfield  {author} {\bibinfo {author} {\bibfnamefont {J.~L.}\ \bibnamefont
  {Egido}},\ }\href {https://doi.org/10.1088/0031-8949/91/7/073003} {\bibfield
  {journal} {\bibinfo  {journal} {Phys. Scr.}\ }\textbf {\bibinfo {volume}
  {91}},\ \bibinfo {pages} {073003} (\bibinfo {year} {2016})}\BibitemShut
  {NoStop}%
\bibitem [{\citenamefont {Robledo}\ \emph {et~al.}(2018)\citenamefont
  {Robledo}, \citenamefont {Rodr{\'{\i}}guez},\ and\ \citenamefont
  {Rodr{\'{\i}}guez-Guzm{\'{a}}n}}]{RR18}%
  \BibitemOpen
  \bibfield  {author} {\bibinfo {author} {\bibfnamefont {L.~M.}\ \bibnamefont
  {Robledo}}, \bibinfo {author} {\bibfnamefont {T.~R.}\ \bibnamefont
  {Rodr{\'{\i}}guez}},\ and\ \bibinfo {author} {\bibfnamefont {R.~R.}\
  \bibnamefont {Rodr{\'{\i}}guez-Guzm{\'{a}}n}},\ }\href
  {https://doi.org/10.1088/1361-6471/aadebd} {\bibfield  {journal} {\bibinfo
  {journal} {J. Phys. G: Nucl. Part.}\ }\textbf {\bibinfo {volume} {46}},\
  \bibinfo {pages} {013001} (\bibinfo {year} {2018})}\BibitemShut {NoStop}%
\bibitem [{\citenamefont {Bender}\ and\ \citenamefont {Heenen}(2003)}]{BH03}%
  \BibitemOpen
  \bibfield  {author} {\bibinfo {author} {\bibfnamefont {M.}~\bibnamefont
  {Bender}}\ and\ \bibinfo {author} {\bibfnamefont {P.-H.}\ \bibnamefont
  {Heenen}},\ }\href
  {https://doi.org/https://doi.org/10.1016/S0375-9474(02)01308-8} {\bibfield
  {journal} {\bibinfo  {journal} {Nucl. Phys. A}\ }\textbf {\bibinfo {volume}
  {713}},\ \bibinfo {pages} {390 } (\bibinfo {year} {2003})}\BibitemShut
  {NoStop}%
\bibitem [{\citenamefont {Duguet}\ \emph {et~al.}(2003)\citenamefont {Duguet},
  \citenamefont {Bender}, \citenamefont {Bonche},\ and\ \citenamefont
  {Heenen}}]{DB03}%
  \BibitemOpen
  \bibfield  {author} {\bibinfo {author} {\bibfnamefont {T.}~\bibnamefont
  {Duguet}}, \bibinfo {author} {\bibfnamefont {M.}~\bibnamefont {Bender}},
  \bibinfo {author} {\bibfnamefont {P.}~\bibnamefont {Bonche}},\ and\ \bibinfo
  {author} {\bibfnamefont {P.-H.}\ \bibnamefont {Heenen}},\ }\href
  {https://doi.org/https://doi.org/10.1016/S0370-2693(03)00330-7} {\bibfield
  {journal} {\bibinfo  {journal} {Phys. Lett. B}\ }\textbf {\bibinfo {volume}
  {559}},\ \bibinfo {pages} {201 } (\bibinfo {year} {2003})}\BibitemShut
  {NoStop}%
\bibitem [{\citenamefont {Bender}\ \emph
  {et~al.}(2003{\natexlab{b}})\citenamefont {Bender}, \citenamefont {Flocard},\
  and\ \citenamefont {Heenen}}]{BF03}%
  \BibitemOpen
  \bibfield  {author} {\bibinfo {author} {\bibfnamefont {M.}~\bibnamefont
  {Bender}}, \bibinfo {author} {\bibfnamefont {H.}~\bibnamefont {Flocard}},\
  and\ \bibinfo {author} {\bibfnamefont {P.~H.}\ \bibnamefont {Heenen}},\
  }\href {https://doi.org/10.1103/PhysRevC.68.044321} {\bibfield  {journal}
  {\bibinfo  {journal} {Phys. Rev. C}\ }\textbf {\bibinfo {volume} {68}},\
  \bibinfo {pages} {044321} (\bibinfo {year} {2003}{\natexlab{b}})}\BibitemShut
  {NoStop}%
\bibitem [{\citenamefont {Rodr\'{\i}guez-Guzm\'an}\ \emph
  {et~al.}(2004)\citenamefont {Rodr\'{\i}guez-Guzm\'an}, \citenamefont
  {Egido},\ and\ \citenamefont {Robledo}}]{RE04}%
  \BibitemOpen
  \bibfield  {author} {\bibinfo {author} {\bibfnamefont {R.~R.}\ \bibnamefont
  {Rodr\'{\i}guez-Guzm\'an}}, \bibinfo {author} {\bibfnamefont {J.~L.}\
  \bibnamefont {Egido}},\ and\ \bibinfo {author} {\bibfnamefont {L.~M.}\
  \bibnamefont {Robledo}},\ }\href {https://doi.org/10.1103/PhysRevC.69.054319}
  {\bibfield  {journal} {\bibinfo  {journal} {Phys. Rev. C}\ }\textbf {\bibinfo
  {volume} {69}},\ \bibinfo {pages} {054319} (\bibinfo {year}
  {2004})}\BibitemShut {NoStop}%
\bibitem [{\citenamefont {Bender}\ \emph {et~al.}(2004)\citenamefont {Bender},
  \citenamefont {Heenen},\ and\ \citenamefont {Bonche}}]{BH04}%
  \BibitemOpen
  \bibfield  {author} {\bibinfo {author} {\bibfnamefont {M.}~\bibnamefont
  {Bender}}, \bibinfo {author} {\bibfnamefont {P.-H.}\ \bibnamefont {Heenen}},\
  and\ \bibinfo {author} {\bibfnamefont {P.}~\bibnamefont {Bonche}},\ }\href
  {https://doi.org/10.1103/PhysRevC.70.054304} {\bibfield  {journal} {\bibinfo
  {journal} {Phys. Rev. C}\ }\textbf {\bibinfo {volume} {70}},\ \bibinfo
  {pages} {054304} (\bibinfo {year} {2004})}\BibitemShut {NoStop}%
\bibitem [{\citenamefont {Shinohara}\ \emph {et~al.}(2006)\citenamefont
  {Shinohara}, \citenamefont {Ohta}, \citenamefont {Nakatsukasa},\ and\
  \citenamefont {Yabana}}]{SO06}%
  \BibitemOpen
  \bibfield  {author} {\bibinfo {author} {\bibfnamefont {S.}~\bibnamefont
  {Shinohara}}, \bibinfo {author} {\bibfnamefont {H.}~\bibnamefont {Ohta}},
  \bibinfo {author} {\bibfnamefont {T.}~\bibnamefont {Nakatsukasa}},\ and\
  \bibinfo {author} {\bibfnamefont {K.}~\bibnamefont {Yabana}},\ }\href
  {https://doi.org/10.1103/PhysRevC.74.054315} {\bibfield  {journal} {\bibinfo
  {journal} {Phys. Rev. C}\ }\textbf {\bibinfo {volume} {74}},\ \bibinfo
  {pages} {054315} (\bibinfo {year} {2006})}\BibitemShut {NoStop}%
\bibitem [{\citenamefont {Rodr\'{\i}guez}\ and\ \citenamefont
  {Egido}(2007)}]{RE07}%
  \BibitemOpen
  \bibfield  {author} {\bibinfo {author} {\bibfnamefont {T.~R.}\ \bibnamefont
  {Rodr\'{\i}guez}}\ and\ \bibinfo {author} {\bibfnamefont {J.~L.}\
  \bibnamefont {Egido}},\ }\href
  {https://doi.org/10.1103/PhysRevLett.99.062501} {\bibfield  {journal}
  {\bibinfo  {journal} {Phys. Rev. Lett.}\ }\textbf {\bibinfo {volume} {99}},\
  \bibinfo {pages} {062501} (\bibinfo {year} {2007})}\BibitemShut {NoStop}%
\bibitem [{\citenamefont {Rodríguez}\ and\ \citenamefont {{Luis
  Egido}}(2008)}]{TR08}%
  \BibitemOpen
  \bibfield  {author} {\bibinfo {author} {\bibfnamefont {T.~R.}\ \bibnamefont
  {Rodríguez}}\ and\ \bibinfo {author} {\bibfnamefont {J.}~\bibnamefont {{Luis
  Egido}}},\ }\href
  {https://doi.org/https://doi.org/10.1016/j.physletb.2008.03.061} {\bibfield
  {journal} {\bibinfo  {journal} {Phys. Lett. B}\ }\textbf {\bibinfo {volume}
  {663}},\ \bibinfo {pages} {49 } (\bibinfo {year} {2008})}\BibitemShut
  {NoStop}%
\bibitem [{\citenamefont {Bender}\ and\ \citenamefont {Heenen}(2008)}]{BH08}%
  \BibitemOpen
  \bibfield  {author} {\bibinfo {author} {\bibfnamefont {M.}~\bibnamefont
  {Bender}}\ and\ \bibinfo {author} {\bibfnamefont {P.-H.}\ \bibnamefont
  {Heenen}},\ }\href {https://doi.org/10.1103/PhysRevC.78.024309} {\bibfield
  {journal} {\bibinfo  {journal} {Phys. Rev. C}\ }\textbf {\bibinfo {volume}
  {78}},\ \bibinfo {pages} {024309} (\bibinfo {year} {2008})}\BibitemShut
  {NoStop}%
\bibitem [{\citenamefont {Yao}\ \emph {et~al.}(2010)\citenamefont {Yao},
  \citenamefont {Meng}, \citenamefont {Ring},\ and\ \citenamefont
  {Vretenar}}]{YM10}%
  \BibitemOpen
  \bibfield  {author} {\bibinfo {author} {\bibfnamefont {J.~M.}\ \bibnamefont
  {Yao}}, \bibinfo {author} {\bibfnamefont {J.}~\bibnamefont {Meng}}, \bibinfo
  {author} {\bibfnamefont {P.}~\bibnamefont {Ring}},\ and\ \bibinfo {author}
  {\bibfnamefont {D.}~\bibnamefont {Vretenar}},\ }\href
  {https://doi.org/10.1103/PhysRevC.81.044311} {\bibfield  {journal} {\bibinfo
  {journal} {Phys. Rev. C}\ }\textbf {\bibinfo {volume} {81}},\ \bibinfo
  {pages} {044311} (\bibinfo {year} {2010})}\BibitemShut {NoStop}%
\bibitem [{\citenamefont {Rodr\'{\i}guez}\ and\ \citenamefont
  {Egido}(2010)}]{RE10}%
  \BibitemOpen
  \bibfield  {author} {\bibinfo {author} {\bibfnamefont {T.~R.}\ \bibnamefont
  {Rodr\'{\i}guez}}\ and\ \bibinfo {author} {\bibfnamefont {J.~L.}\
  \bibnamefont {Egido}},\ }\href {https://doi.org/10.1103/PhysRevC.81.064323}
  {\bibfield  {journal} {\bibinfo  {journal} {Phys. Rev. C}\ }\textbf {\bibinfo
  {volume} {81}},\ \bibinfo {pages} {064323} (\bibinfo {year}
  {2010})}\BibitemShut {NoStop}%
\bibitem [{\citenamefont {Rodr\'{\i}guez}\ and\ \citenamefont
  {Egido}(2011{\natexlab{a}})}]{RT11}%
  \BibitemOpen
  \bibfield  {author} {\bibinfo {author} {\bibfnamefont {T.~R.}\ \bibnamefont
  {Rodr\'{\i}guez}}\ and\ \bibinfo {author} {\bibfnamefont {J.~L.}\
  \bibnamefont {Egido}},\ }\href {https://doi.org/10.1103/PhysRevC.84.051307}
  {\bibfield  {journal} {\bibinfo  {journal} {Phys. Rev. C}\ }\textbf {\bibinfo
  {volume} {84}},\ \bibinfo {pages} {051307} (\bibinfo {year}
  {2011}{\natexlab{a}})}\BibitemShut {NoStop}%
\bibitem [{\citenamefont {Rodríguez}\ and\ \citenamefont
  {Egido}(2011)}]{TR11}%
  \BibitemOpen
  \bibfield  {author} {\bibinfo {author} {\bibfnamefont {T.~R.}\ \bibnamefont
  {Rodríguez}}\ and\ \bibinfo {author} {\bibfnamefont {J.~L.}\ \bibnamefont
  {Egido}},\ }\href
  {https://doi.org/https://doi.org/10.1016/j.physletb.2011.10.003} {\bibfield
  {journal} {\bibinfo  {journal} {Phys. Lett. B}\ }\textbf {\bibinfo {volume}
  {705}},\ \bibinfo {pages} {255 } (\bibinfo {year} {2011})}\BibitemShut
  {NoStop}%
\bibitem [{\citenamefont {Rodr\'{\i}guez}\ and\ \citenamefont
  {Egido}(2011{\natexlab{b}})}]{RE11}%
  \BibitemOpen
  \bibfield  {author} {\bibinfo {author} {\bibfnamefont {T.~R.}\ \bibnamefont
  {Rodr\'{\i}guez}}\ and\ \bibinfo {author} {\bibfnamefont {J.~L.}\
  \bibnamefont {Egido}},\ }\href {https://doi.org/10.1103/PhysRevC.84.051307}
  {\bibfield  {journal} {\bibinfo  {journal} {Phys. Rev. C}\ }\textbf {\bibinfo
  {volume} {84}},\ \bibinfo {pages} {051307} (\bibinfo {year}
  {2011}{\natexlab{b}})}\BibitemShut {NoStop}%
\bibitem [{\citenamefont {Yao}\ \emph {et~al.}(2013)\citenamefont {Yao},
  \citenamefont {Bender},\ and\ \citenamefont {Heenen}}]{YB13}%
  \BibitemOpen
  \bibfield  {author} {\bibinfo {author} {\bibfnamefont {J.~M.}\ \bibnamefont
  {Yao}}, \bibinfo {author} {\bibfnamefont {M.}~\bibnamefont {Bender}},\ and\
  \bibinfo {author} {\bibfnamefont {P.-H.}\ \bibnamefont {Heenen}},\ }\href
  {https://doi.org/10.1103/PhysRevC.87.034322} {\bibfield  {journal} {\bibinfo
  {journal} {Phys. Rev. C}\ }\textbf {\bibinfo {volume} {87}},\ \bibinfo
  {pages} {034322} (\bibinfo {year} {2013})}\BibitemShut {NoStop}%
\bibitem [{\citenamefont {Fukuoka}\ \emph {et~al.}(2013)\citenamefont
  {Fukuoka}, \citenamefont {Shinohara}, \citenamefont {Funaki}, \citenamefont
  {Nakatsukasa},\ and\ \citenamefont {Yabana}}]{FS13}%
  \BibitemOpen
  \bibfield  {author} {\bibinfo {author} {\bibfnamefont {Y.}~\bibnamefont
  {Fukuoka}}, \bibinfo {author} {\bibfnamefont {S.}~\bibnamefont {Shinohara}},
  \bibinfo {author} {\bibfnamefont {Y.}~\bibnamefont {Funaki}}, \bibinfo
  {author} {\bibfnamefont {T.}~\bibnamefont {Nakatsukasa}},\ and\ \bibinfo
  {author} {\bibfnamefont {K.}~\bibnamefont {Yabana}},\ }\href
  {https://doi.org/10.1103/PhysRevC.88.014321} {\bibfield  {journal} {\bibinfo
  {journal} {Phys. Rev. C}\ }\textbf {\bibinfo {volume} {88}},\ \bibinfo
  {pages} {014321} (\bibinfo {year} {2013})}\BibitemShut {NoStop}%
\bibitem [{\citenamefont {Bally}\ \emph {et~al.}(2014)\citenamefont {Bally},
  \citenamefont {Avez}, \citenamefont {Bender},\ and\ \citenamefont
  {Heenen}}]{BA14}%
  \BibitemOpen
  \bibfield  {author} {\bibinfo {author} {\bibfnamefont {B.}~\bibnamefont
  {Bally}}, \bibinfo {author} {\bibfnamefont {B.}~\bibnamefont {Avez}},
  \bibinfo {author} {\bibfnamefont {M.}~\bibnamefont {Bender}},\ and\ \bibinfo
  {author} {\bibfnamefont {P.-H.}\ \bibnamefont {Heenen}},\ }\href
  {https://doi.org/10.1103/PhysRevLett.113.162501} {\bibfield  {journal}
  {\bibinfo  {journal} {Phys. Rev. Lett.}\ }\textbf {\bibinfo {volume} {113}},\
  \bibinfo {pages} {162501} (\bibinfo {year} {2014})}\BibitemShut {NoStop}%
\bibitem [{\citenamefont {Yao}\ \emph {et~al.}(2014)\citenamefont {Yao},
  \citenamefont {Hagino}, \citenamefont {Li}, \citenamefont {Meng},\ and\
  \citenamefont {Ring}}]{Yao2014}%
  \BibitemOpen
  \bibfield  {author} {\bibinfo {author} {\bibfnamefont {J.~M.}\ \bibnamefont
  {Yao}}, \bibinfo {author} {\bibfnamefont {K.}~\bibnamefont {Hagino}},
  \bibinfo {author} {\bibfnamefont {Z.~P.}\ \bibnamefont {Li}}, \bibinfo
  {author} {\bibfnamefont {J.}~\bibnamefont {Meng}},\ and\ \bibinfo {author}
  {\bibfnamefont {P.}~\bibnamefont {Ring}},\ }\href
  {https://doi.org/10.1103/PhysRevC.89.054306} {\bibfield  {journal} {\bibinfo
  {journal} {Phys. Rev. C}\ }\textbf {\bibinfo {volume} {89}},\ \bibinfo
  {pages} {054306} (\bibinfo {year} {2014})}\BibitemShut {NoStop}%
\bibitem [{\citenamefont {Yao}\ \emph {et~al.}(2015)\citenamefont {Yao},
  \citenamefont {Zhou},\ and\ \citenamefont {Li}}]{YZ15}%
  \BibitemOpen
  \bibfield  {author} {\bibinfo {author} {\bibfnamefont {J.~M.}\ \bibnamefont
  {Yao}}, \bibinfo {author} {\bibfnamefont {E.~F.}\ \bibnamefont {Zhou}},\ and\
  \bibinfo {author} {\bibfnamefont {Z.~P.}\ \bibnamefont {Li}},\ }\href
  {https://doi.org/10.1103/PhysRevC.92.041304} {\bibfield  {journal} {\bibinfo
  {journal} {Phys. Rev. C}\ }\textbf {\bibinfo {volume} {92}},\ \bibinfo
  {pages} {041304} (\bibinfo {year} {2015})}\BibitemShut {NoStop}%
\bibitem [{\citenamefont {Egido}\ and\ \citenamefont {Jungclaus}(2020)}]{EJ20}%
  \BibitemOpen
  \bibfield  {author} {\bibinfo {author} {\bibfnamefont {J.~L.}\ \bibnamefont
  {Egido}}\ and\ \bibinfo {author} {\bibfnamefont {A.}~\bibnamefont
  {Jungclaus}},\ }\href {https://doi.org/10.1103/PhysRevLett.125.192504}
  {\bibfield  {journal} {\bibinfo  {journal} {Phys. Rev. Lett.}\ }\textbf
  {\bibinfo {volume} {125}},\ \bibinfo {pages} {192504} (\bibinfo {year}
  {2020})}\BibitemShut {NoStop}%
\bibitem [{\citenamefont {Ring}\ and\ \citenamefont
  {Schuck}(1980)}]{Ring_Schuck}%
  \BibitemOpen
  \bibfield  {author} {\bibinfo {author} {\bibfnamefont {P.}~\bibnamefont
  {Ring}}\ and\ \bibinfo {author} {\bibfnamefont {P.}~\bibnamefont {Schuck}},\
  }\href@noop {} {\emph {\bibinfo {title} {The nuclear many-body problem}}}\
  (\bibinfo  {publisher} {Springer-Verlag},\ \bibinfo {address} {New York},\
  \bibinfo {year} {1980})\BibitemShut {NoStop}%
\bibitem [{\citenamefont {Peierls}\ and\ \citenamefont
  {Thouless}(1962)}]{PT62}%
  \BibitemOpen
  \bibfield  {author} {\bibinfo {author} {\bibfnamefont {P.}~\bibnamefont
  {Peierls}}\ and\ \bibinfo {author} {\bibfnamefont {D.}~\bibnamefont
  {Thouless}},\ }\href
  {https://doi.org/https://doi.org/10.1016/0029-5582(62)91025-8} {\bibfield
  {journal} {\bibinfo  {journal} {Nucl. Phys.}\ }\textbf {\bibinfo {volume}
  {38}},\ \bibinfo {pages} {154 } (\bibinfo {year} {1962})}\BibitemShut
  {NoStop}%
\bibitem [{\citenamefont {Borrajo}\ \emph {et~al.}(2015)\citenamefont
  {Borrajo}, \citenamefont {Rodríguez},\ and\ \citenamefont {{Luis
  Egido}}}]{BR15}%
  \BibitemOpen
  \bibfield  {author} {\bibinfo {author} {\bibfnamefont {M.}~\bibnamefont
  {Borrajo}}, \bibinfo {author} {\bibfnamefont {T.~R.}\ \bibnamefont
  {Rodríguez}},\ and\ \bibinfo {author} {\bibfnamefont {J.}~\bibnamefont
  {{Luis Egido}}},\ }\href
  {https://doi.org/https://doi.org/10.1016/j.physletb.2015.05.030} {\bibfield
  {journal} {\bibinfo  {journal} {Phys. Lett. B}\ }\textbf {\bibinfo {volume}
  {746}},\ \bibinfo {pages} {341 } (\bibinfo {year} {2015})}\BibitemShut
  {NoStop}%
\bibitem [{\citenamefont {Egido}\ \emph {et~al.}(2016)\citenamefont {Egido},
  \citenamefont {Borrajo},\ and\ \citenamefont {Rodr\'{\i}guez}}]{EB16}%
  \BibitemOpen
  \bibfield  {author} {\bibinfo {author} {\bibfnamefont {J.~L.}\ \bibnamefont
  {Egido}}, \bibinfo {author} {\bibfnamefont {M.}~\bibnamefont {Borrajo}},\
  and\ \bibinfo {author} {\bibfnamefont {T.~R.}\ \bibnamefont
  {Rodr\'{\i}guez}},\ }\href {https://doi.org/10.1103/PhysRevLett.116.052502}
  {\bibfield  {journal} {\bibinfo  {journal} {Phys. Rev. Lett.}\ }\textbf
  {\bibinfo {volume} {116}},\ \bibinfo {pages} {052502} (\bibinfo {year}
  {2016})}\BibitemShut {NoStop}%
\bibitem [{\citenamefont {Shimada}\ \emph {et~al.}(2015)\citenamefont
  {Shimada}, \citenamefont {Tagami},\ and\ \citenamefont {Shimizu}}]{ST15}%
  \BibitemOpen
  \bibfield  {author} {\bibinfo {author} {\bibfnamefont {M.}~\bibnamefont
  {Shimada}}, \bibinfo {author} {\bibfnamefont {S.}~\bibnamefont {Tagami}},\
  and\ \bibinfo {author} {\bibfnamefont {Y.~R.}\ \bibnamefont {Shimizu}},\
  }\href@noop {} {\bibfield  {journal} {\bibinfo  {journal} {Prog. Theo. Exp.
  Phys}\ }\textbf {\bibinfo {volume} {2015}},\ \bibinfo {pages} {063D02}
  (\bibinfo {year} {2015})}\BibitemShut {NoStop}%
\bibitem [{\citenamefont {Shimada}\ \emph {et~al.}(2016)\citenamefont
  {Shimada}, \citenamefont {Tagami},\ and\ \citenamefont
  {Shimizu}}]{Shimada2016}%
  \BibitemOpen
  \bibfield  {author} {\bibinfo {author} {\bibfnamefont {M.}~\bibnamefont
  {Shimada}}, \bibinfo {author} {\bibfnamefont {S.}~\bibnamefont {Tagami}},\
  and\ \bibinfo {author} {\bibfnamefont {Y.~R.}\ \bibnamefont {Shimizu}},\
  }\href {https://doi.org/10.1103/PhysRevC.93.044317} {\bibfield  {journal}
  {\bibinfo  {journal} {Phys. Rev. C}\ }\textbf {\bibinfo {volume} {93}},\
  \bibinfo {pages} {044317} (\bibinfo {year} {2016})}\BibitemShut {NoStop}%
\bibitem [{\citenamefont {Ushitani}\ \emph {et~al.}(2019)\citenamefont
  {Ushitani}, \citenamefont {Tagami},\ and\ \citenamefont
  {Shimizu}}]{Ushitani2019}%
  \BibitemOpen
  \bibfield  {author} {\bibinfo {author} {\bibfnamefont {M.}~\bibnamefont
  {Ushitani}}, \bibinfo {author} {\bibfnamefont {S.}~\bibnamefont {Tagami}},\
  and\ \bibinfo {author} {\bibfnamefont {Y.~R.}\ \bibnamefont {Shimizu}},\
  }\href {https://doi.org/10.1103/PhysRevC.99.064328} {\bibfield  {journal}
  {\bibinfo  {journal} {Phys. Rev. C}\ }\textbf {\bibinfo {volume} {99}},\
  \bibinfo {pages} {064328} (\bibinfo {year} {2019})}\BibitemShut {NoStop}%
\bibitem [{\citenamefont {Jancovici}\ and\ \citenamefont
  {Schiff}(1964)}]{jan64}%
  \BibitemOpen
  \bibfield  {author} {\bibinfo {author} {\bibfnamefont {B.}~\bibnamefont
  {Jancovici}}\ and\ \bibinfo {author} {\bibfnamefont {D.}~\bibnamefont
  {Schiff}},\ }\href
  {https://doi.org/https://doi.org/10.1016/0029-5582(64)90578-4} {\bibfield
  {journal} {\bibinfo  {journal} {Nucl. Phys.}\ }\textbf {\bibinfo {volume}
  {58}},\ \bibinfo {pages} {678 } (\bibinfo {year} {1964})}\BibitemShut
  {NoStop}%
\bibitem [{\citenamefont {Brink}\ and\ \citenamefont {Weiguny}(1968)}]{BW68}%
  \BibitemOpen
  \bibfield  {author} {\bibinfo {author} {\bibfnamefont {D.}~\bibnamefont
  {Brink}}\ and\ \bibinfo {author} {\bibfnamefont {A.}~\bibnamefont
  {Weiguny}},\ }\href
  {https://doi.org/https://doi.org/10.1016/0375-9474(68)90059-6} {\bibfield
  {journal} {\bibinfo  {journal} {Nucl. Phys. A}\ }\textbf {\bibinfo {volume}
  {120}},\ \bibinfo {pages} {59 } (\bibinfo {year} {1968})}\BibitemShut
  {NoStop}%
\bibitem [{\citenamefont {Goeke}\ and\ \citenamefont {Reinhard}(1978)}]{GR78}%
  \BibitemOpen
  \bibfield  {author} {\bibinfo {author} {\bibfnamefont {K.}~\bibnamefont
  {Goeke}}\ and\ \bibinfo {author} {\bibfnamefont {P.-G.}\ \bibnamefont
  {Reinhard}},\ }\href
  {https://doi.org/https://doi.org/10.1016/S0003-4916(78)80003-7} {\bibfield
  {journal} {\bibinfo  {journal} {Ann. Phys. (N. Y.)}\ }\textbf {\bibinfo
  {volume} {112}},\ \bibinfo {pages} {328 } (\bibinfo {year}
  {1978})}\BibitemShut {NoStop}%
\bibitem [{\citenamefont {Reinhard}\ and\ \citenamefont
  {Goeke}(1978{\natexlab{a}})}]{RG78a}%
  \BibitemOpen
  \bibfield  {author} {\bibinfo {author} {\bibfnamefont {P.~G.}\ \bibnamefont
  {Reinhard}}\ and\ \bibinfo {author} {\bibfnamefont {K.}~\bibnamefont
  {Goeke}},\ }\href {https://doi.org/10.1103/PhysRevC.17.1249} {\bibfield
  {journal} {\bibinfo  {journal} {Phys. Rev. C}\ }\textbf {\bibinfo {volume}
  {17}},\ \bibinfo {pages} {1249} (\bibinfo {year}
  {1978}{\natexlab{a}})}\BibitemShut {NoStop}%
\bibitem [{\citenamefont {Reinhard}\ and\ \citenamefont
  {Goeke}(1978{\natexlab{b}})}]{RG78b}%
  \BibitemOpen
  \bibfield  {author} {\bibinfo {author} {\bibfnamefont {P.~G.}\ \bibnamefont
  {Reinhard}}\ and\ \bibinfo {author} {\bibfnamefont {K.}~\bibnamefont
  {Goeke}},\ }\href {https://doi.org/10.1088/0305-4616/4/9/006} {\bibfield
  {journal} {\bibinfo  {journal} {J. Phys. G}\ }\textbf {\bibinfo {volume}
  {4}},\ \bibinfo {pages} {L245} (\bibinfo {year}
  {1978}{\natexlab{b}})}\BibitemShut {NoStop}%
\bibitem [{\citenamefont {Reinhard}\ and\ \citenamefont
  {Goeke}(1978{\natexlab{c}})}]{RG78c}%
  \BibitemOpen
  \bibfield  {author} {\bibinfo {author} {\bibfnamefont {P.-G.}\ \bibnamefont
  {Reinhard}}\ and\ \bibinfo {author} {\bibfnamefont {K.}~\bibnamefont
  {Goeke}},\ }\href
  {https://doi.org/https://doi.org/10.1016/0375-9474(78)90579-1} {\bibfield
  {journal} {\bibinfo  {journal} {Nucl. Phys. A}\ }\textbf {\bibinfo {volume}
  {312}},\ \bibinfo {pages} {121 } (\bibinfo {year}
  {1978}{\natexlab{c}})}\BibitemShut {NoStop}%
\bibitem [{\citenamefont {Goeke}\ and\ \citenamefont {Reinhard}(1980)}]{GR80}%
  \BibitemOpen
  \bibfield  {author} {\bibinfo {author} {\bibfnamefont {K.}~\bibnamefont
  {Goeke}}\ and\ \bibinfo {author} {\bibfnamefont {P.-G.}\ \bibnamefont
  {Reinhard}},\ }\href
  {https://doi.org/https://doi.org/10.1016/0003-4916(80)90210-9} {\bibfield
  {journal} {\bibinfo  {journal} {Ann. Phys. (N. Y.)}\ }\textbf {\bibinfo
  {volume} {124}},\ \bibinfo {pages} {249 } (\bibinfo {year}
  {1980})}\BibitemShut {NoStop}%
\bibitem [{\citenamefont {{López Vaquero}}\ \emph {et~al.}(2011)\citenamefont
  {{López Vaquero}}, \citenamefont {Rodriguez},\ and\ \citenamefont
  {Egido}}]{VR11}%
  \BibitemOpen
  \bibfield  {author} {\bibinfo {author} {\bibfnamefont {N.}~\bibnamefont
  {{López Vaquero}}}, \bibinfo {author} {\bibfnamefont {T.~R.}\ \bibnamefont
  {Rodriguez}},\ and\ \bibinfo {author} {\bibfnamefont {J.~L.}\ \bibnamefont
  {Egido}},\ }\href
  {https://doi.org/https://doi.org/10.1016/j.physletb.2011.09.073} {\bibfield
  {journal} {\bibinfo  {journal} {Phys. Lett. B}\ }\textbf {\bibinfo {volume}
  {704}},\ \bibinfo {pages} {520 } (\bibinfo {year} {2011})}\BibitemShut
  {NoStop}%
\bibitem [{\citenamefont {Broglia}\ and\ \citenamefont
  {Zelevinsky}(2013)}]{FYNB}%
  \BibitemOpen
  \bibfield  {author} {\bibinfo {author} {\bibfnamefont {R.~A.}\ \bibnamefont
  {Broglia}}\ and\ \bibinfo {author} {\bibfnamefont {V.}~\bibnamefont
  {Zelevinsky}},\ }\href {https://doi.org/10.1142/8526} {\emph {\bibinfo
  {title} {Fifty Years of Nuclear BCS}}}\ (\bibinfo  {publisher} {World
  Scientific},\ \bibinfo {year} {2013})\BibitemShut {NoStop}%
\bibitem [{\citenamefont {Vaquero}\ \emph {et~al.}(2013)\citenamefont
  {Vaquero}, \citenamefont {Rodr\'{\i}guez},\ and\ \citenamefont
  {Egido}}]{Vaquero2013}%
  \BibitemOpen
  \bibfield  {author} {\bibinfo {author} {\bibfnamefont {N.~L.}\ \bibnamefont
  {Vaquero}}, \bibinfo {author} {\bibfnamefont {T.~R.}\ \bibnamefont
  {Rodr\'{\i}guez}},\ and\ \bibinfo {author} {\bibfnamefont {J.~L.}\
  \bibnamefont {Egido}},\ }\href
  {https://doi.org/10.1103/PhysRevLett.111.142501} {\bibfield  {journal}
  {\bibinfo  {journal} {Phys. Rev. Lett.}\ }\textbf {\bibinfo {volume} {111}},\
  \bibinfo {pages} {142501} (\bibinfo {year} {2013})}\BibitemShut {NoStop}%
\bibitem [{\citenamefont {Giuliani}\ \emph {et~al.}(2014)\citenamefont
  {Giuliani}, \citenamefont {Robledo},\ and\ \citenamefont
  {Rodr\'{\i}guez-Guzm\'an}}]{Giuliani2014}%
  \BibitemOpen
  \bibfield  {author} {\bibinfo {author} {\bibfnamefont {S.~A.}\ \bibnamefont
  {Giuliani}}, \bibinfo {author} {\bibfnamefont {L.~M.}\ \bibnamefont
  {Robledo}},\ and\ \bibinfo {author} {\bibfnamefont {R.}~\bibnamefont
  {Rodr\'{\i}guez-Guzm\'an}},\ }\href
  {https://doi.org/10.1103/PhysRevC.90.054311} {\bibfield  {journal} {\bibinfo
  {journal} {Phys. Rev. C}\ }\textbf {\bibinfo {volume} {90}},\ \bibinfo
  {pages} {054311} (\bibinfo {year} {2014})}\BibitemShut {NoStop}%
\bibitem [{\citenamefont {Erler}\ \emph {et~al.}(2012)\citenamefont {Erler},
  \citenamefont {Birge}, \citenamefont {Kortelainen}, \citenamefont
  {Nazarewicz}, \citenamefont {Olsen}, \citenamefont {Perhac},\ and\
  \citenamefont {Stoitsov}}]{erl12}%
  \BibitemOpen
  \bibfield  {author} {\bibinfo {author} {\bibfnamefont {J.}~\bibnamefont
  {Erler}}, \bibinfo {author} {\bibfnamefont {N.}~\bibnamefont {Birge}},
  \bibinfo {author} {\bibfnamefont {M.}~\bibnamefont {Kortelainen}}, \bibinfo
  {author} {\bibfnamefont {W.}~\bibnamefont {Nazarewicz}}, \bibinfo {author}
  {\bibfnamefont {E.}~\bibnamefont {Olsen}}, \bibinfo {author} {\bibfnamefont
  {A.~M.}\ \bibnamefont {Perhac}},\ and\ \bibinfo {author} {\bibfnamefont
  {M.}~\bibnamefont {Stoitsov}},\ }\href {https://doi.org/10.1038/nature11188}
  {\bibfield  {journal} {\bibinfo  {journal} {Nature}\ }\textbf {\bibinfo
  {volume} {486}},\ \bibinfo {pages} {509} (\bibinfo {year}
  {2012})}\BibitemShut {NoStop}%
\bibitem [{\citenamefont {Beiner}\ \emph {et~al.}(1975)\citenamefont {Beiner},
  \citenamefont {Flocard}, \citenamefont {{Van Giai}},\ and\ \citenamefont
  {Quentin}}]{BF75}%
  \BibitemOpen
  \bibfield  {author} {\bibinfo {author} {\bibfnamefont {M.}~\bibnamefont
  {Beiner}}, \bibinfo {author} {\bibfnamefont {H.}~\bibnamefont {Flocard}},
  \bibinfo {author} {\bibfnamefont {N.}~\bibnamefont {{Van Giai}}},\ and\
  \bibinfo {author} {\bibfnamefont {P.}~\bibnamefont {Quentin}},\ }\href
  {https://doi.org/https://doi.org/10.1016/0375-9474(75)90338-3} {\bibfield
  {journal} {\bibinfo  {journal} {Nucl. Phys. A}\ }\textbf {\bibinfo {volume}
  {238}},\ \bibinfo {pages} {29 } (\bibinfo {year} {1975})}\BibitemShut
  {NoStop}%
\bibitem [{\citenamefont {Dobaczewski}\ \emph {et~al.}(2007)\citenamefont
  {Dobaczewski}, \citenamefont {Stoitsov}, \citenamefont {Nazarewicz},\ and\
  \citenamefont {Reinhard}}]{DS07}%
  \BibitemOpen
  \bibfield  {author} {\bibinfo {author} {\bibfnamefont {J.}~\bibnamefont
  {Dobaczewski}}, \bibinfo {author} {\bibfnamefont {M.~V.}\ \bibnamefont
  {Stoitsov}}, \bibinfo {author} {\bibfnamefont {W.}~\bibnamefont
  {Nazarewicz}},\ and\ \bibinfo {author} {\bibfnamefont {P.-G.}\ \bibnamefont
  {Reinhard}},\ }\href {https://doi.org/10.1103/PhysRevC.76.054315} {\bibfield
  {journal} {\bibinfo  {journal} {Phys. Rev. C}\ }\textbf {\bibinfo {volume}
  {76}},\ \bibinfo {pages} {054315} (\bibinfo {year} {2007})}\BibitemShut
  {NoStop}%
\bibitem [{\citenamefont {Bonche}\ \emph {et~al.}(1990)\citenamefont {Bonche},
  \citenamefont {Dobaczewski}, \citenamefont {Flocard}, \citenamefont
  {Heenen},\ and\ \citenamefont {Meyer}}]{BD90}%
  \BibitemOpen
  \bibfield  {author} {\bibinfo {author} {\bibfnamefont {P.}~\bibnamefont
  {Bonche}}, \bibinfo {author} {\bibfnamefont {J.}~\bibnamefont {Dobaczewski}},
  \bibinfo {author} {\bibfnamefont {H.}~\bibnamefont {Flocard}}, \bibinfo
  {author} {\bibfnamefont {P.-H.}\ \bibnamefont {Heenen}},\ and\ \bibinfo
  {author} {\bibfnamefont {J.}~\bibnamefont {Meyer}},\ }\href
  {https://doi.org/https://doi.org/10.1016/0375-9474(90)90062-Q} {\bibfield
  {journal} {\bibinfo  {journal} {Nucl. Phys. A}\ }\textbf {\bibinfo {volume}
  {510}},\ \bibinfo {pages} {466 } (\bibinfo {year} {1990})}\BibitemShut
  {NoStop}%
\bibitem [{LAP()}]{LAPACK}%
  \BibitemOpen
  \href@noop {} {\bibinfo {title} {{LAPACK}}},\ \bibinfo {howpublished}
  {\url{http://www.netlib.org/lapack/}}\BibitemShut {NoStop}%
\bibitem [{Note1()}]{Note1}%
  \BibitemOpen
  \bibinfo {note} {For $^{16}$O, the problem of overcompliteness is found to be
  severe, and we chose $\lambda _{\protect \mathrm {cut}}=8.0\times 10^{-2}$,
  which is determined from the eigenvalue distribution of the overlap
  kernel.}\BibitemShut {Stop}%
\bibitem [{\citenamefont {Sheikh}\ and\ \citenamefont
  {Ring}(2000)}]{Sheikh2000}%
  \BibitemOpen
  \bibfield  {author} {\bibinfo {author} {\bibfnamefont {J.}~\bibnamefont
  {Sheikh}}\ and\ \bibinfo {author} {\bibfnamefont {P.}~\bibnamefont {Ring}},\
  }\href {https://doi.org/10.1016/S0375-9474(99)00424-8} {\bibfield  {journal}
  {\bibinfo  {journal} {Nucl. Phys. A}\ }\textbf {\bibinfo {volume} {665}},\
  \bibinfo {pages} {71} (\bibinfo {year} {2000})}\BibitemShut {NoStop}%
\bibitem [{\citenamefont {Sheikh}\ \emph {et~al.}(2002)\citenamefont {Sheikh},
  \citenamefont {Ring}, \citenamefont {Lopes},\ and\ \citenamefont
  {Rossignoli}}]{Sheikh2002}%
  \BibitemOpen
  \bibfield  {author} {\bibinfo {author} {\bibfnamefont {J.~A.}\ \bibnamefont
  {Sheikh}}, \bibinfo {author} {\bibfnamefont {P.}~\bibnamefont {Ring}},
  \bibinfo {author} {\bibfnamefont {E.}~\bibnamefont {Lopes}},\ and\ \bibinfo
  {author} {\bibfnamefont {R.}~\bibnamefont {Rossignoli}},\ }\href
  {https://doi.org/10.1103/PhysRevC.66.044318} {\bibfield  {journal} {\bibinfo
  {journal} {Phys. Rev. C}\ }\textbf {\bibinfo {volume} {66}},\ \bibinfo
  {pages} {044318} (\bibinfo {year} {2002})}\BibitemShut {NoStop}%
\bibitem [{\citenamefont {Stoitsov}\ \emph {et~al.}(2007)\citenamefont
  {Stoitsov}, \citenamefont {Dobaczewski}, \citenamefont {Kirchner},
  \citenamefont {Nazarewicz},\ and\ \citenamefont {Terasaki}}]{Stoitsov2007}%
  \BibitemOpen
  \bibfield  {author} {\bibinfo {author} {\bibfnamefont {M.~V.}\ \bibnamefont
  {Stoitsov}}, \bibinfo {author} {\bibfnamefont {J.}~\bibnamefont
  {Dobaczewski}}, \bibinfo {author} {\bibfnamefont {R.}~\bibnamefont
  {Kirchner}}, \bibinfo {author} {\bibfnamefont {W.}~\bibnamefont
  {Nazarewicz}},\ and\ \bibinfo {author} {\bibfnamefont {J.}~\bibnamefont
  {Terasaki}},\ }\href {https://doi.org/10.1103/PhysRevC.76.014308} {\bibfield
  {journal} {\bibinfo  {journal} {Phys. Rev. C}\ }\textbf {\bibinfo {volume}
  {76}},\ \bibinfo {pages} {014308} (\bibinfo {year} {2007})}\BibitemShut
  {NoStop}%
\bibitem [{\citenamefont {Duguet}\ \emph {et~al.}(2009)\citenamefont {Duguet},
  \citenamefont {Bender}, \citenamefont {Bennaceur}, \citenamefont {Lacroix},\
  and\ \citenamefont {Lesinski}}]{Duguet2009}%
  \BibitemOpen
  \bibfield  {author} {\bibinfo {author} {\bibfnamefont {T.}~\bibnamefont
  {Duguet}}, \bibinfo {author} {\bibfnamefont {M.}~\bibnamefont {Bender}},
  \bibinfo {author} {\bibfnamefont {K.}~\bibnamefont {Bennaceur}}, \bibinfo
  {author} {\bibfnamefont {D.}~\bibnamefont {Lacroix}},\ and\ \bibinfo {author}
  {\bibfnamefont {T.}~\bibnamefont {Lesinski}},\ }\href
  {https://doi.org/10.1103/PhysRevC.79.044320} {\bibfield  {journal} {\bibinfo
  {journal} {Phys. Rev. C}\ }\textbf {\bibinfo {volume} {79}},\ \bibinfo
  {pages} {044320} (\bibinfo {year} {2009})}\BibitemShut {NoStop}%
\bibitem [{\citenamefont {Hupin}\ and\ \citenamefont
  {Lacroix}(2012)}]{Hupin2012}%
  \BibitemOpen
  \bibfield  {author} {\bibinfo {author} {\bibfnamefont {G.}~\bibnamefont
  {Hupin}}\ and\ \bibinfo {author} {\bibfnamefont {D.}~\bibnamefont
  {Lacroix}},\ }\href {https://doi.org/10.1103/PhysRevC.86.024309} {\bibfield
  {journal} {\bibinfo  {journal} {Phys. Rev. C}\ }\textbf {\bibinfo {volume}
  {86}},\ \bibinfo {pages} {024309} (\bibinfo {year} {2012})}\BibitemShut
  {NoStop}%
\bibitem [{\citenamefont {Lipkin}(1960)}]{L60}%
  \BibitemOpen
  \bibfield  {author} {\bibinfo {author} {\bibfnamefont {H.~J.}\ \bibnamefont
  {Lipkin}},\ }\href
  {https://doi.org/https://doi.org/10.1016/0003-4916(60)90032-4} {\bibfield
  {journal} {\bibinfo  {journal} {Ann. Phys. (N. Y.)}\ }\textbf {\bibinfo
  {volume} {9}},\ \bibinfo {pages} {272 } (\bibinfo {year} {1960})}\BibitemShut
  {NoStop}%
\bibitem [{\citenamefont {Nogami}(1964)}]{N64}%
  \BibitemOpen
  \bibfield  {author} {\bibinfo {author} {\bibfnamefont {Y.}~\bibnamefont
  {Nogami}},\ }\href {https://doi.org/10.1103/PhysRev.134.B313} {\bibfield
  {journal} {\bibinfo  {journal} {Phys. Rev.}\ }\textbf {\bibinfo {volume}
  {134}},\ \bibinfo {pages} {B313} (\bibinfo {year} {1964})}\BibitemShut
  {NoStop}%
\bibitem [{\citenamefont {Zheng}\ \emph {et~al.}(1992)\citenamefont {Zheng},
  \citenamefont {Sprung},\ and\ \citenamefont {Flocard}}]{ZS92}%
  \BibitemOpen
  \bibfield  {author} {\bibinfo {author} {\bibfnamefont {D.~C.}\ \bibnamefont
  {Zheng}}, \bibinfo {author} {\bibfnamefont {D.~W.~L.}\ \bibnamefont
  {Sprung}},\ and\ \bibinfo {author} {\bibfnamefont {H.}~\bibnamefont
  {Flocard}},\ }\href {https://doi.org/10.1103/PhysRevC.46.1355} {\bibfield
  {journal} {\bibinfo  {journal} {Phys. Rev. C}\ }\textbf {\bibinfo {volume}
  {46}},\ \bibinfo {pages} {1355} (\bibinfo {year} {1992})}\BibitemShut
  {NoStop}%
\bibitem [{\citenamefont {Dobaczewski}\ and\ \citenamefont
  {Nazarewicz}(1993)}]{DN93}%
  \BibitemOpen
  \bibfield  {author} {\bibinfo {author} {\bibfnamefont {J.}~\bibnamefont
  {Dobaczewski}}\ and\ \bibinfo {author} {\bibfnamefont {W.}~\bibnamefont
  {Nazarewicz}},\ }\href {https://doi.org/10.1103/PhysRevC.47.2418} {\bibfield
  {journal} {\bibinfo  {journal} {Phys. Rev. C}\ }\textbf {\bibinfo {volume}
  {47}},\ \bibinfo {pages} {2418} (\bibinfo {year} {1993})}\BibitemShut
  {NoStop}%
\bibitem [{\citenamefont {Hagino}\ and\ \citenamefont {Bertsch}(2000)}]{HB00}%
  \BibitemOpen
  \bibfield  {author} {\bibinfo {author} {\bibfnamefont {K.}~\bibnamefont
  {Hagino}}\ and\ \bibinfo {author} {\bibfnamefont {G.}~\bibnamefont
  {Bertsch}},\ }\href
  {https://doi.org/https://doi.org/10.1016/S0375-9474(00)00343-2} {\bibfield
  {journal} {\bibinfo  {journal} {Nucl. Phys. A}\ }\textbf {\bibinfo {volume}
  {679}},\ \bibinfo {pages} {163 } (\bibinfo {year} {2000})}\BibitemShut
  {NoStop}%
\bibitem [{\citenamefont {Hagino}\ \emph {et~al.}(2002)\citenamefont {Hagino},
  \citenamefont {Reinhard},\ and\ \citenamefont {Bertsch}}]{Hagino2002}%
  \BibitemOpen
  \bibfield  {author} {\bibinfo {author} {\bibfnamefont {K.}~\bibnamefont
  {Hagino}}, \bibinfo {author} {\bibfnamefont {P.-G.}\ \bibnamefont
  {Reinhard}},\ and\ \bibinfo {author} {\bibfnamefont {G.~F.}\ \bibnamefont
  {Bertsch}},\ }\href {https://doi.org/10.1103/PhysRevC.65.064320} {\bibfield
  {journal} {\bibinfo  {journal} {Phys. Rev. C}\ }\textbf {\bibinfo {volume}
  {65}},\ \bibinfo {pages} {064320} (\bibinfo {year} {2002})}\BibitemShut
  {NoStop}%
\bibitem [{\citenamefont {Mottelson}\ and\ \citenamefont
  {Valatin}(1960)}]{MV60}%
  \BibitemOpen
  \bibfield  {author} {\bibinfo {author} {\bibfnamefont {B.~R.}\ \bibnamefont
  {Mottelson}}\ and\ \bibinfo {author} {\bibfnamefont {J.~G.}\ \bibnamefont
  {Valatin}},\ }\href {https://doi.org/10.1103/PhysRevLett.5.511} {\bibfield
  {journal} {\bibinfo  {journal} {Phys. Rev. Lett.}\ }\textbf {\bibinfo
  {volume} {5}},\ \bibinfo {pages} {511} (\bibinfo {year} {1960})}\BibitemShut
  {NoStop}%
\bibitem [{\citenamefont {Egido}\ \emph {et~al.}(1985)\citenamefont {Egido},
  \citenamefont {Ring}, \citenamefont {Iwasaki},\ and\ \citenamefont
  {Mang}}]{ER85}%
  \BibitemOpen
  \bibfield  {author} {\bibinfo {author} {\bibfnamefont {J.}~\bibnamefont
  {Egido}}, \bibinfo {author} {\bibfnamefont {P.}~\bibnamefont {Ring}},
  \bibinfo {author} {\bibfnamefont {S.}~\bibnamefont {Iwasaki}},\ and\ \bibinfo
  {author} {\bibfnamefont {H.}~\bibnamefont {Mang}},\ }\href
  {https://doi.org/https://doi.org/10.1016/0370-2693(85)91555-2} {\bibfield
  {journal} {\bibinfo  {journal} {Phys. Lett. B}\ }\textbf {\bibinfo {volume}
  {154}},\ \bibinfo {pages} {1 } (\bibinfo {year} {1985})}\BibitemShut
  {NoStop}%
\bibitem [{\citenamefont {Shimizu}\ and\ \citenamefont {Broglia}(1990)}]{SB90}%
  \BibitemOpen
  \bibfield  {author} {\bibinfo {author} {\bibfnamefont {Y.}~\bibnamefont
  {Shimizu}}\ and\ \bibinfo {author} {\bibfnamefont {R.}~\bibnamefont
  {Broglia}},\ }\href
  {https://doi.org/https://doi.org/10.1016/0375-9474(90)90321-C} {\bibfield
  {journal} {\bibinfo  {journal} {Nucl. Phys. A}\ }\textbf {\bibinfo {volume}
  {515}},\ \bibinfo {pages} {38 } (\bibinfo {year} {1990})}\BibitemShut
  {NoStop}%
\bibitem [{\citenamefont {Wigner}(1937)}]{W37}%
  \BibitemOpen
  \bibfield  {author} {\bibinfo {author} {\bibfnamefont {E.}~\bibnamefont
  {Wigner}},\ }\href {https://doi.org/10.1103/PhysRev.51.106} {\bibfield
  {journal} {\bibinfo  {journal} {Phys. Rev.}\ }\textbf {\bibinfo {volume}
  {51}},\ \bibinfo {pages} {106} (\bibinfo {year} {1937})}\BibitemShut
  {NoStop}%
\bibitem [{\citenamefont {Van~Isacker}\ \emph {et~al.}(1995)\citenamefont
  {Van~Isacker}, \citenamefont {Warner},\ and\ \citenamefont {Brenner}}]{IW95}%
  \BibitemOpen
  \bibfield  {author} {\bibinfo {author} {\bibfnamefont {P.}~\bibnamefont
  {Van~Isacker}}, \bibinfo {author} {\bibfnamefont {D.~D.}\ \bibnamefont
  {Warner}},\ and\ \bibinfo {author} {\bibfnamefont {D.~S.}\ \bibnamefont
  {Brenner}},\ }\href {https://doi.org/10.1103/PhysRevLett.74.4607} {\bibfield
  {journal} {\bibinfo  {journal} {Phys. Rev. Lett.}\ }\textbf {\bibinfo
  {volume} {74}},\ \bibinfo {pages} {4607} (\bibinfo {year}
  {1995})}\BibitemShut {NoStop}%
\bibitem [{\citenamefont {Satuła}\ and\ \citenamefont
  {Wyss}(1997)}]{Satula1997}%
  \BibitemOpen
  \bibfield  {author} {\bibinfo {author} {\bibfnamefont {W.}~\bibnamefont
  {Satuła}}\ and\ \bibinfo {author} {\bibfnamefont {R.}~\bibnamefont {Wyss}},\
  }\href {https://doi.org/https://doi.org/10.1016/S0370-2693(96)01603-6}
  {\bibfield  {journal} {\bibinfo  {journal} {Phys. Lett. B}\ }\textbf
  {\bibinfo {volume} {393}},\ \bibinfo {pages} {1} (\bibinfo {year}
  {1997})}\BibitemShut {NoStop}%
\bibitem [{\citenamefont {Satuła}\ \emph {et~al.}(1997)\citenamefont
  {Satuła}, \citenamefont {Dean}, \citenamefont {Gary}, \citenamefont
  {Mizutori},\ and\ \citenamefont {Nazarewicz}}]{Satula1997-2}%
  \BibitemOpen
  \bibfield  {author} {\bibinfo {author} {\bibfnamefont {W.}~\bibnamefont
  {Satuła}}, \bibinfo {author} {\bibfnamefont {D.}~\bibnamefont {Dean}},
  \bibinfo {author} {\bibfnamefont {J.}~\bibnamefont {Gary}}, \bibinfo {author}
  {\bibfnamefont {S.}~\bibnamefont {Mizutori}},\ and\ \bibinfo {author}
  {\bibfnamefont {W.}~\bibnamefont {Nazarewicz}},\ }\href
  {https://doi.org/https://doi.org/10.1016/S0370-2693(97)00711-9} {\bibfield
  {journal} {\bibinfo  {journal} {Phys. Lett. B}\ }\textbf {\bibinfo {volume}
  {407}},\ \bibinfo {pages} {103} (\bibinfo {year} {1997})}\BibitemShut
  {NoStop}%
\bibitem [{\citenamefont {Bender}\ \emph {et~al.}(2020)\citenamefont {Bender},
  \citenamefont {Bernard}, \citenamefont {Bertsch}, \citenamefont {Chiba},
  \citenamefont {Dobaczewski}, \citenamefont {Dubray}, \citenamefont
  {Giuliani}, \citenamefont {Hagino}, \citenamefont {Lacroix}, \citenamefont
  {Li}, \citenamefont {Magierski}, \citenamefont {Maruhn}, \citenamefont
  {Nazarewicz}, \citenamefont {Pei}, \citenamefont {P{\'{e}}ru}, \citenamefont
  {Pillet}, \citenamefont {Randrup}, \citenamefont {Regnier}, \citenamefont
  {Reinhard}, \citenamefont {Robledo}, \citenamefont {Ryssens}, \citenamefont
  {Sadhukhan}, \citenamefont {Scamps}, \citenamefont {Schunck}, \citenamefont
  {Simenel}, \citenamefont {Skalski}, \citenamefont {Stetcu}, \citenamefont
  {Stevenson}, \citenamefont {Umar}, \citenamefont {Verriere}, \citenamefont
  {Vretenar}, \citenamefont {Warda},\ and\ \citenamefont
  {{\AA}berg}}]{Bender2020}%
  \BibitemOpen
  \bibfield  {author} {\bibinfo {author} {\bibfnamefont {M.}~\bibnamefont
  {Bender}}, \bibinfo {author} {\bibfnamefont {R.}~\bibnamefont {Bernard}},
  \bibinfo {author} {\bibfnamefont {G.}~\bibnamefont {Bertsch}}, \bibinfo
  {author} {\bibfnamefont {S.}~\bibnamefont {Chiba}}, \bibinfo {author}
  {\bibfnamefont {J.}~\bibnamefont {Dobaczewski}}, \bibinfo {author}
  {\bibfnamefont {N.}~\bibnamefont {Dubray}}, \bibinfo {author} {\bibfnamefont
  {S.~A.}\ \bibnamefont {Giuliani}}, \bibinfo {author} {\bibfnamefont
  {K.}~\bibnamefont {Hagino}}, \bibinfo {author} {\bibfnamefont
  {D.}~\bibnamefont {Lacroix}}, \bibinfo {author} {\bibfnamefont
  {Z.}~\bibnamefont {Li}}, \bibinfo {author} {\bibfnamefont {P.}~\bibnamefont
  {Magierski}}, \bibinfo {author} {\bibfnamefont {J.}~\bibnamefont {Maruhn}},
  \bibinfo {author} {\bibfnamefont {W.}~\bibnamefont {Nazarewicz}}, \bibinfo
  {author} {\bibfnamefont {J.}~\bibnamefont {Pei}}, \bibinfo {author}
  {\bibfnamefont {S.}~\bibnamefont {P{\'{e}}ru}}, \bibinfo {author}
  {\bibfnamefont {N.}~\bibnamefont {Pillet}}, \bibinfo {author} {\bibfnamefont
  {J.}~\bibnamefont {Randrup}}, \bibinfo {author} {\bibfnamefont
  {D.}~\bibnamefont {Regnier}}, \bibinfo {author} {\bibfnamefont {P.-G.}\
  \bibnamefont {Reinhard}}, \bibinfo {author} {\bibfnamefont {L.~M.}\
  \bibnamefont {Robledo}}, \bibinfo {author} {\bibfnamefont {W.}~\bibnamefont
  {Ryssens}}, \bibinfo {author} {\bibfnamefont {J.}~\bibnamefont {Sadhukhan}},
  \bibinfo {author} {\bibfnamefont {G.}~\bibnamefont {Scamps}}, \bibinfo
  {author} {\bibfnamefont {N.}~\bibnamefont {Schunck}}, \bibinfo {author}
  {\bibfnamefont {C.}~\bibnamefont {Simenel}}, \bibinfo {author} {\bibfnamefont
  {J.}~\bibnamefont {Skalski}}, \bibinfo {author} {\bibfnamefont
  {I.}~\bibnamefont {Stetcu}}, \bibinfo {author} {\bibfnamefont
  {P.}~\bibnamefont {Stevenson}}, \bibinfo {author} {\bibfnamefont
  {S.}~\bibnamefont {Umar}}, \bibinfo {author} {\bibfnamefont {M.}~\bibnamefont
  {Verriere}}, \bibinfo {author} {\bibfnamefont {D.}~\bibnamefont {Vretenar}},
  \bibinfo {author} {\bibfnamefont {M.}~\bibnamefont {Warda}},\ and\ \bibinfo
  {author} {\bibfnamefont {S.}~\bibnamefont {{\AA}berg}},\ }\href
  {https://doi.org/10.1088/1361-6471/abab4f} {\bibfield  {journal} {\bibinfo
  {journal} {J. Phys. G: Nucl. Part.}\ }\textbf {\bibinfo {volume} {47}},\
  \bibinfo {pages} {113002} (\bibinfo {year} {2020})}\BibitemShut {NoStop}%
\end{thebibliography}%

\end{document}